\documentclass[preprint]{aastex62}
\usepackage{natbib}
\usepackage{graphicx}
\usepackage{epstopdf}
\usepackage{mathtools}
\usepackage{bm}
\newcommand{\be}{\begin{equation}}
\newcommand{\ee}{\end{equation} }
\newcommand{\ba}{\begin{eqnarray}}
\newcommand{\ea}{\end{eqnarray}}

\newcommand{\nn}{\mbox{} \nonumber \\ \mbox{} }
\newcommand{\kB}{k_{\rm B}}

\shortauthors{Thompson}	
\received{2018 October 31}
\published{2019 March 20}
\begin{document}
\title{Impulsive Electromagnetic Emission near a Black Hole}
\author{Christopher Thompson}
\affil{Canadian Institute for Theoretical Astrophysics, 60 St. George St., Toronto, ON M5S 3H8, Canada.}		

\begin{abstract}
The electromagnetic signature of a point explosion near a Kerr black hole (BH) is evaluated.   The first repetitions 
produced by gravitational lensing are not periodic in time;  periodicity emerges only as the result
of multiple circuits of the prograde and retrograde light rings and is accompanied by exponential dimming.
Gravitational focusing creates a sequence of concentrated caustic features and biases the detection of a
repeating source toward alignment of the BH spin with the plane of the sky.   We consider the polarization
pattern in the case of emission by the Lorentz upboosting and reflection of a magnetic field near the explosion site.
Then the polarized fraction of the detected pulse approaches unity, and rays propagating near the equatorial plane maintain
a consistent polarization direction.  Near a slowly accreting supermassive BH (SMBH), additional repetitions are caused by
reflection off annular fragments of an orbiting disk that has passed through an ionization instability.   These 
results are applied to the repeating fast radio burst (FRB) source 121102, giving a concrete and predictive example of how FRB detectability
may be biased by lensing.    A gravitational lensing delay of 10-30 s, and reflection delay up to $\sim 10^4$ s, are 
found for emission near the innermost stable circular orbit 
of a $3\times 10^5\,M_\odot$ SMBH; these effects combine to produce interesting correlations
between delay time and burst fluence.  A similar repetitive pulse envelope could be seen in the gravitational wave signal
produced by a collision between compact stars near a SMBH.
\end{abstract}

\section{Introduction}\label{s:intro}

This article considers the impulsive injection of electromagnetic fields within a small volume next to a spinning black
hole (BH).  The radiation pattern at infinity shows a discrete time structure and forms a sequence of caustic
features, which we describe along with the polarization pattern.  Our focus is on emission by particles (or collisions
between them) in orbits confined to the equatorial plane of the hole.   

The signal from a continuously emitting point source orbiting a BH was first considered by \cite{cb72}
and has been analyzed subsequently by a number of authors:  see \cite{bl05}, \cite{dn17}, and \cite{gralla18}, and
references therein.  Continuous emission viewed at a large distance from the BH is strongly modulated at the orbital
period of the source,  whereas, as we show, the signal 
of impulsive energy release develops periodicity only with some delay and with exponentially diminishing intensity.

In effect, we have mapped out the light-cone component of the electromagnetic Green function of a Kerr BH on the sphere
at infinity, although only
for a particular location and velocity of the source.  The analogous problem of impulsive scalar field emission near 
a Schwarzschild BH has been treated by \cite{zg12} using a numerical relativity code, revealing a caustic 
structure identical to the one uncovered here in the zero-spin case.  The simple ray-tracing method adopted here is more effective
at extracting narrow temporal features in the extended signal and can easily incorporate propagation effects such as
reflection.  A much more elaborate caustic structure emerges when the BH spins, which complicates a direct comparison
with the known spin-1 quasi-normal modes of a Kerr BH.  The mapping of this caustic structure onto the observer's
plane for a continuously emitting point source of variable displacement from the BH was performed by \cite{rb94}
and \cite{bozza08}.

When a BH is a source of repeated point electromagnetic bursts, gravitational focusing creates a strong detection bias
favoring the alignment of the BH spin with the plane of the sky.  The repetitive pattern described here 
would also be imposed on the gravitational wave signal produced by a collision between neutron stars or
stellar-mass BHs near a supermassive black hole (SMBH).  

Part of the motivation for this study came from our proposal \citep{thompson17a, thompson17b} that fast 
radio bursts (FRBs) signal collisions between macroscopic dipolar dark matter particles and that repeating FRBs
are powered by a small fraction of these particles that are captured by SMBHs into a dynamically cold
ring close to the innermost stable circular orbit (ISCO).  A few predictions of that model (strong linear polarization, 
extreme Faraday rotation, and strong high-frequency emission)
are consistent with recent data collected from the only currently known repeating FRB 121102 \citep{gajjar18,michilli18}.
Measurements of time delays between successive bursts indicate a dearth of repetitions shorter than $\sim 10$ s,
which is consistent with gravitational lensing by a $\sim 3\times 10^5\,M_\odot$ BH.

Electromagnetic bursts could be emitted over a range of frequencies, indeed with the emission peaking above 10-100 GHz
in the aforementioned model.  The escape of lower-frequency gigahertz waves from close to a SMBH requires the electron density
to be low, less than $\sim 10^7$ cm$^{-3}$ for FRB 121102 after taking into account the feedback of the strong
electromagnetic wave on ambient electrons \citep{thompson17b}.  In that situation, a thin disk component of the accretion
flow onto a SMBH will have passed through an ionization transition and frozen out, becoming susceptible to warping and
breaking \citep{ns03,nealon15}.  Especially for the BH spin orientation favored by lensing detection bias, specular
reflection off this orbiting material is an additional source of repetitions over $10^{2-3}$ s intervals.  

The plan of this article is as follows.  Section \ref{s:ray} describes our handling of
point emission and ray propagation in a Kerr spacetime.  A Monte
Carlo method is used to map the trapped and escaping rays and to determine the pattern of delayed pulses.
As is shown in Section \ref{s:raych},
pulse arrival times and fluences can be calculated over much longer delays in the case of equatorial rays by a simple
application of Raychaudhuri's equations.  Section \ref{s:bias} shows how the detection volume of the pulses depends on
the orientation of the observer relative to the equatorial plane of the BH.  In Section \ref{s:reflect} we turn to consider
hydrodynamical effects, in particular the mass profile of a quiescent accretion disk that passes through an ionization
transition, and the characteristic displacement of warps and annular fragments of such a disk from the SMBH.  Section
\ref{s:frb} applies our results to the repeating FRB source 121102, and Section \ref{s:summary} summarizes.
The appendices give further details and tests of our calculations, in particular demonstrating the equivalence of a global
Monte Carlo calculation with the integration of the ray expansion along equatorial geodesics.

\section{Ray Emission and Propagation}\label{s:ray}

We consider the isotropic emission of photons (rays), concentrated at a single time, 
near a BH of mass $M$ and angular momentum $Ma$.  The Kerr line element,
expressed in Boyer-Lindquist (B-L) coordinates,\footnote{Unless otherwise indicated, we use geometrical units 
in which $c = G = 1$.} is
\ba\label{eq:kerr}
ds^2 &=& \widetilde g_{tt}dt^2 + g_{\phi\phi}[d(\phi-\omega t)]^2 + g_{rr}dr^2 + g_{\theta\theta}d\theta^2\nn
     &=& -{\Sigma \Delta\over A}dt^2 + {A\over\Sigma}\sin^2\theta[d(\phi-\omega t)]^2 + {\Sigma\over\Delta}dr^2 + 
      \Sigma d\theta^2.
\ea
Here 
\be \label{eq:omega}
\omega = -{g_{t\phi}\over g_{\phi\phi}} = {2M ar\over A}
\ee
is the angular frequency of a zero angular momentum observer (ZAMO),
$\widetilde g_{tt} \equiv g_{tt} - \omega^2g_{\phi\phi}$, and
$\Sigma(r,\theta) = r^2 + a^2\cos^2\theta$, $\Delta(r) = r^2 -2Mr + a^2$, and 
$A(r,\theta) = (r^2 + a^2)^2 - a^2\Delta\sin^2\theta$.  The asymptotic direction of escaping rays is recorded
in colatitude and azimuth $(\theta,\phi)$.

Emission is assumed to take place in the frame of a massive particle moving in a prograde
circular orbit of angular velocity
\be\label{eq:omegac}
\omega_c = {(Mr)^{1/2}\over r^2 + a(Mr)^{1/2}}.
\ee
We particularly explore emission at the ISCO, whose radius depends on the spin and mass of the hole \citep{bpt72}
\ba
r_{\rm isco} &=&  M\left\{3 + Z_2 - [(3-Z_1)(3+Z_1+2Z_2)]^{1/2}\right\};\nn
Z_1 &\equiv& 1 + (1-a^2/M^2)^{1/3}[(1+a/M)^{1/3} + (1-a/M)^{1/3}];\nn
Z_2 &\equiv& (3a^2/M^2 + Z_1^2)^{1/3}.
\ea
For example, repeating FRB emission could arise from collisions between superconducting dipoles that gradually
lose orbital energy by hydromagnetic dissipation, ending up in a dynamically cold ring near the ISCO of a SMBH
\citep{thompson17a}.

The photon wavevector in this emission frame has components $k^a_{\rm em}$ given by
\be\label{eq:ki}
k^{(1)}_{\rm em} = k^{(0)}_{\rm em} \sin\theta_{\rm em}\cos\phi_{\rm em}; \quad\quad k^{(2)}_{\rm em} = -k^{(0)}_{\rm em}\cos\theta_{\rm em}; \quad\quad 
k^{(3)}_{\rm em} = k^{(0)}_{\rm em} \sin\theta_{\rm em}\sin\phi_{\rm em}.
\ee
Here $\theta_{\rm em}$ and $\phi_{\rm em}$ are angles on a small sphere surrounding the emission point, with
$\theta_{\rm em} = 0$ corresponding to the direction $-\hat\theta$, $\theta_{\rm em} = \pi/2$, $\phi_{\rm em} = 0$ to the 
direction $\hat r$, and $\theta_{\rm em} = \pi/2$, $\phi_{\rm em} = \pi/2$ to the direction $\hat\phi$.   The ray tangent
vector is related to the emission-frame wavevector by $dx^\mu/d\lambda = e^\mu_a k^a_{\rm em}$, where $\lambda$ is
the affine parameter, and the tetrad
\ba\label{eq:tet}
e^\mu_{(0)}\partial_\mu &=&  {\gamma_{\rm em}\over (-\widetilde g_{tt})^{1/2}}\partial_t + 
{\gamma_{\rm em}(\beta_{\rm em} + \beta_\omega)\over g_{\phi\phi}^{1/2}}\partial_\phi; \nn
e^r_{(1)} &=& g_{rr}^{-1/2}; \quad\quad e^\theta_{(2)} = g_{\theta\theta}^{-1/2}\nn
e^\mu_{(3)}\partial_\mu &=& {\gamma_{\rm em}\beta_{\rm em}\over (-\widetilde g_{tt})^{1/2}}\partial_t + 
{\gamma_{\rm em}(1 + \beta_{\rm em}\beta_\omega)\over g_{\phi\phi}^{1/2}}\partial_\phi.
\ea
Here
\be\label{eq:betadef}
\beta_\omega = {g_{\phi\phi}^{1/2}\over (-\widetilde g_{tt})^{1/2}}\omega
\ee
and
\be
\beta_{\rm em} = { g_{\phi\phi}^{1/2}\over (-\widetilde g_{tt})^{1/2}}(\omega_c - \omega);\quad\quad
\gamma_{\rm em} = (1-\beta_{\rm em}^2)^{-1/2}
\ee
is the boost connecting the ZAMO frame to the emission frame.

The electric vector of the emitted ray is determined by a simple geometrical model, which is based on the emission
process described by \cite{rees77} and \cite{blandford77} in the context of BH evaporation, and is also
one of the processes that can power radio emission from a tiny electromagnetic explosion \citep{thompson17b}.
Here an electrically conducting, spherical shell expands outward radially from the emission point.  The ambient 
magnetic field (with direction $\hat B_{\rm ex}$ in the emission frame) is observed as a counter-propagating 
electromagnetic wave in the rest frame of the shell and, in the emission frame, is upboosted into a superluminal wave 
with unit electric vector $\hat E = -\hat R_{\rm em}\times \hat B_{\rm ex}/|\hat R_{\rm em}\times\hat B_{\rm ex}|$, 
where $\hat R_{\rm em}$ is the radial unit
vector in the direction of expansion.  We choose for calculational purposes a toroidal ambient magnetic field, 
$\hat B_{\rm ex} = \pm\hat\phi$.  The initial B-L frame polarization is then given by $\varepsilon^\mu_{\rm em} = 
\pm e^\mu_a (\hat R_{\rm em} \times \hat\phi)^a / |\hat R_{\rm em}\times\hat\phi|$.  

Ray propagation is handled by evolving the geodesic equation
\be
{d^2x^\mu\over d\lambda^2} = - \Gamma^\mu_{\alpha\beta}{dx^\alpha\over d\lambda} {dx^\beta\over d\lambda}.
\ee
The orbital energy and angular momentum integrals are
\ba\label{eq:integrals}
{\cal E} &=& -g_{t\mu}{dx^\mu\over d\lambda} = -\widetilde g_{tt} {dt\over d\lambda} + \omega {\cal L}_z; \nn
{\cal L}_z &=& g_{\phi\mu}{dx^\mu\over d\lambda} = g_{\phi\phi}\left({d\phi\over d\lambda} -\omega{dt\over d\lambda}\right),
\ea
and are expressed in terms of emission-frame quantities by
\ba\label{eq:integrals2}
{\cal E} &=& (-\widetilde g_{tt})^{1/2} \gamma_{\rm em}\left[1+\beta_{\rm em}\beta_\omega + (\beta_{\rm em} + \beta_\omega)\hat k^{(3)}\right]
k_{\rm em}^{(0)}; \nn
{\cal L}_z &=& g_{\phi\phi}^{1/2} \gamma_{\rm em}(\beta_{\rm em} + \hat k^{(3)})k_{\rm em}^{(0)}.
\ea
Although the ray trajectory can also be obtained by quadrature, by combining 
${\cal E}$ and ${\cal L}_z$ with the Carter constant (see \citealt{fn98} and Appendix \ref{s:test}), 
a direct integration of the geodesic equation has several advantages:  it allows one easily to include rays with 
turning points, to resolve
delayed narrow pulses and construct maps, and to analyze correlations between variables (e.g. between winding number or
time delay and the minimum approach to the BH).   The integrations are easily accomplished to high accuracy 
($\sim 10^{-12}$ or better) using a package such as DLSODE \citep{hindmarsh83}.

The polarization 4-vector also evolves by parallel transport,
\be
{d\varepsilon^\mu\over d\lambda} = -\Gamma^\mu_{\alpha\beta}{dx^\alpha\over d\lambda} \varepsilon^\beta.
\ee
One finds that $\varepsilon^\mu$ develops a longitudinal component parallel to $k^\mu$, but the physical 
polarization is easily read off at large radius by truncating to the $\phi$ and $\theta$ components.

\begin{figure}
\epsscale{0.9}
\plotone{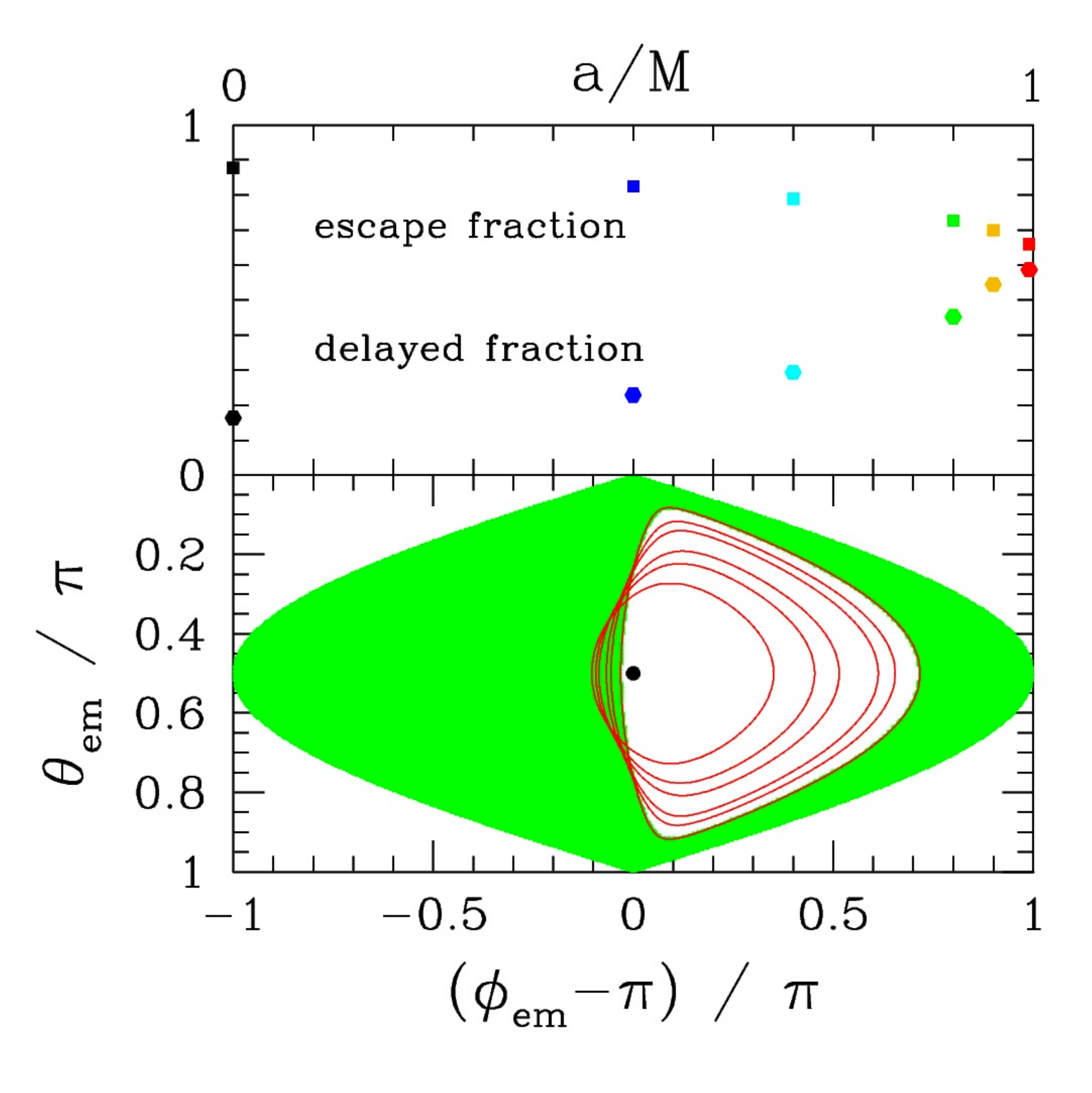}
\caption{Bottom panel:  trapped (white) and escaping (green) rays emitted by a spherical point flash from the
ISCO of a Kerr BH with dimensionless spin $a/M = 0.99$, as mapped onto a small sphere co-orbiting with the point
source (Equation (\ref{eq:ki})).  Black dot denotes the direction of the center of the hole, with rays to the right
being emitted against the orbital motion.  Red lines show the boundary between trapped and escaping rays for a
sequence of BH spins, $a/M = 0$, 0.5, 0.7, 0.9, 0.95, 0.99.  Top panel: fraction of the escaping rays appearing in
delayed pulses (hexagonal dots), and fraction of all rays escaping to infinity (square dots).}
\label{fig:escape_map}
\end{figure}

\subsection{Absorption vs. Escape}\label{s:escape}

The rays that are absorbed by the BH can be mapped out on a small sphere surrounding the emission
point, defined by the angles (\ref{eq:ki}).  The result is shown in Figure \ref{fig:escape_map} for a
range of $a/M$.  The angular zone of accreted rays grows in size with rising $a/M$ as a result of the strengthening
of spacetime curvature at the ISCO and the increasing apparent angular size of the hole.  Rays propagating to the
poles ($\theta_{\rm em} = 0, \pi$) escape to infinity.

The net fraction of escaping rays, under the assumption of isotropic emission, is shown also as a function of 
the BH spin in Figure \ref{fig:escape_map}.  In the case of anisotropic emission, this result would still 
apply after averaging over many explosions with random orientation.

\begin{figure}
\epsscale{0.55}
\plotone{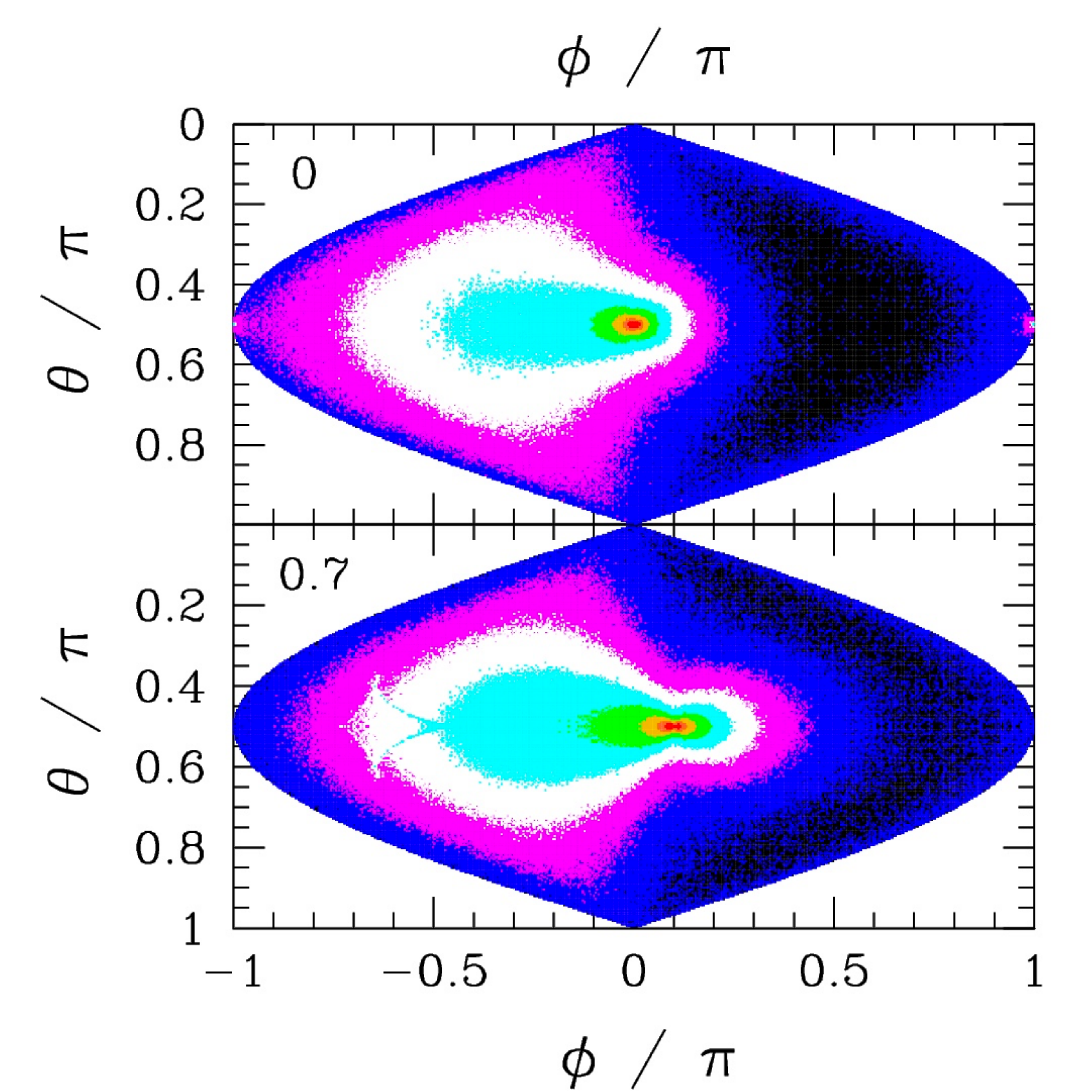}
\vskip .2in
\plotone{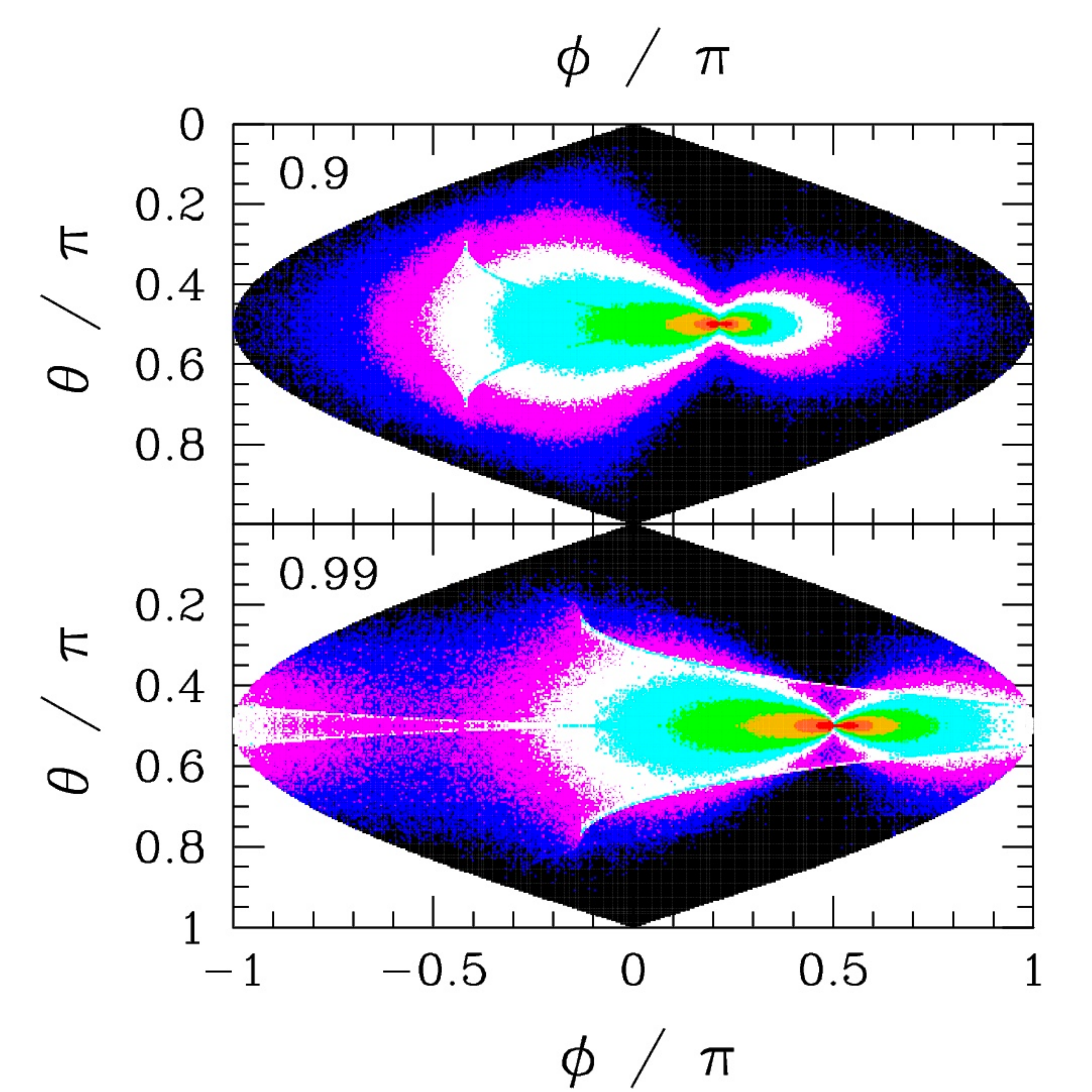}
\vskip .2in
\caption{Ray fluence (defined by Equation (\ref{eq:fluence})) on the sphere at infinity,
corresponding to point emission in the frame of a massive test particle on the ISCO of
a Kerr BH.  The BH spin increases from top to bottom:  $a/M = 0$, 0.7, 0.9, 0.99.  Colors mark zones 
with fluence separated by powers of 1.7 (1.8 bottom panel), increasing from black to blue, mauve, white, cyan, green, 
gold, and red.  The point $(\theta=0,\,\phi=0)$ is antipodal to the position of the explosion.
Strong peak near the center is produced by gravitational lensing, and the broad concentration
to its left is caused by the motion of the emission frame.
Projection is equal area, with horizontal lines marking constant rotational colatitude $\theta$.
These maps are obtained by a Monte Carlo procedure involving $10^{23}$ trial rays, whose asymptotic direction
is recorded in pixels of size ($2^{-8}\pi,\, 2^{-8}\pi/\sin\theta$).}
\label{fig:flux_map}
\vskip 0.2in
\end{figure}

\begin{figure}
\epsscale{0.8}
\vskip -0.3in
\plotone{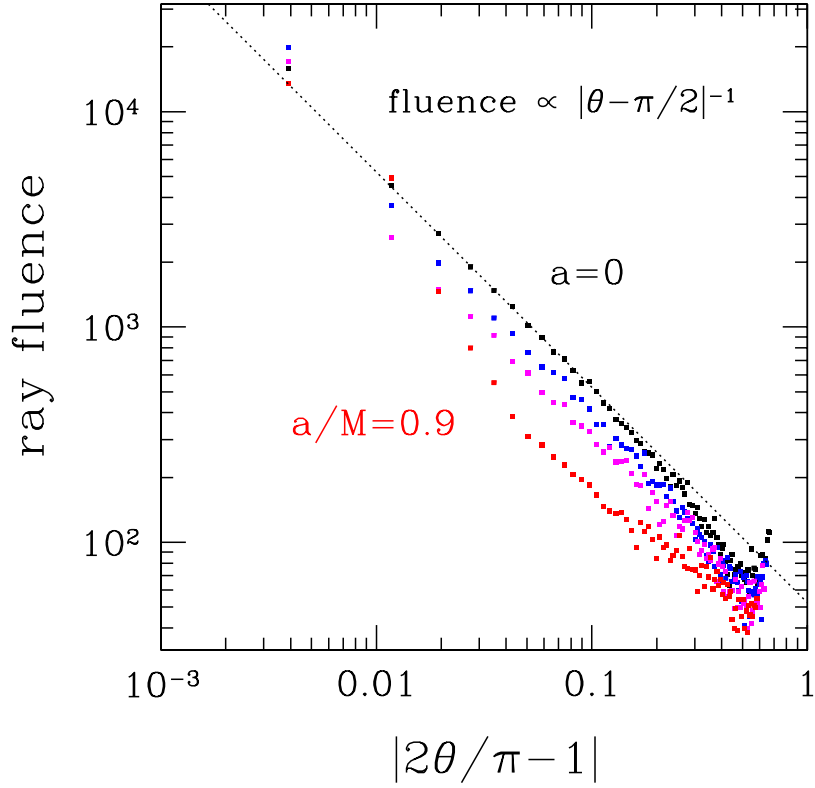}
\vskip -1in
\caption{Dependence of ray fluence on latitude above and below the equatorial caustic on the sphere at infinity,
as see in Figure \ref{fig:flux_map}.  BH spins range from $a = 0$ (black points) to $a/M = 0.9$ (red points).
Angular dependence is $|\theta-\pi/2|^{-1}$ in the simplest case of a non-spinning BH (dotted line).}
\label{fig:flux_lat}
\vskip 0.2in
\end{figure}

\begin{figure}
\epsscale{1.05}
\plottwo{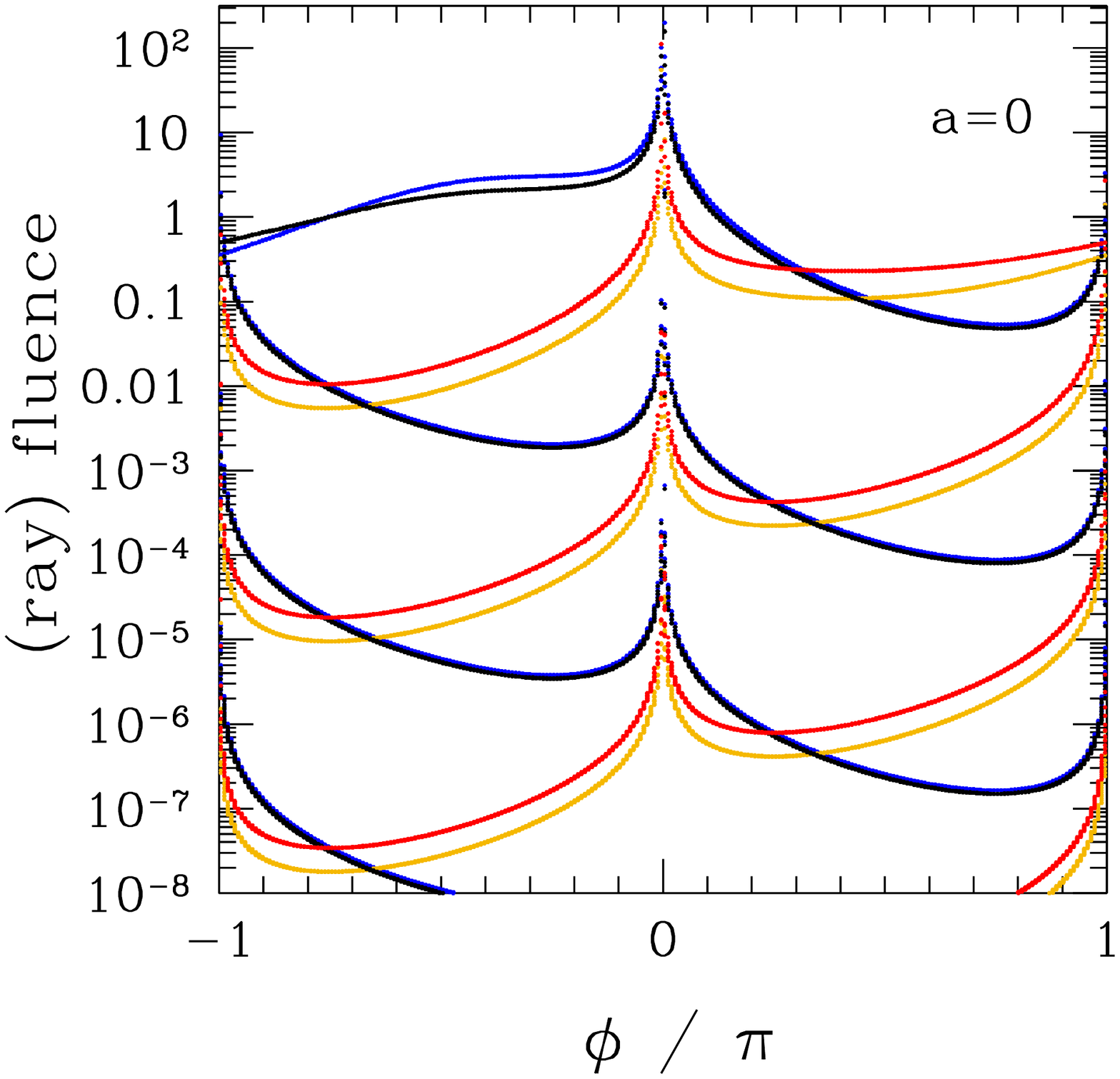}{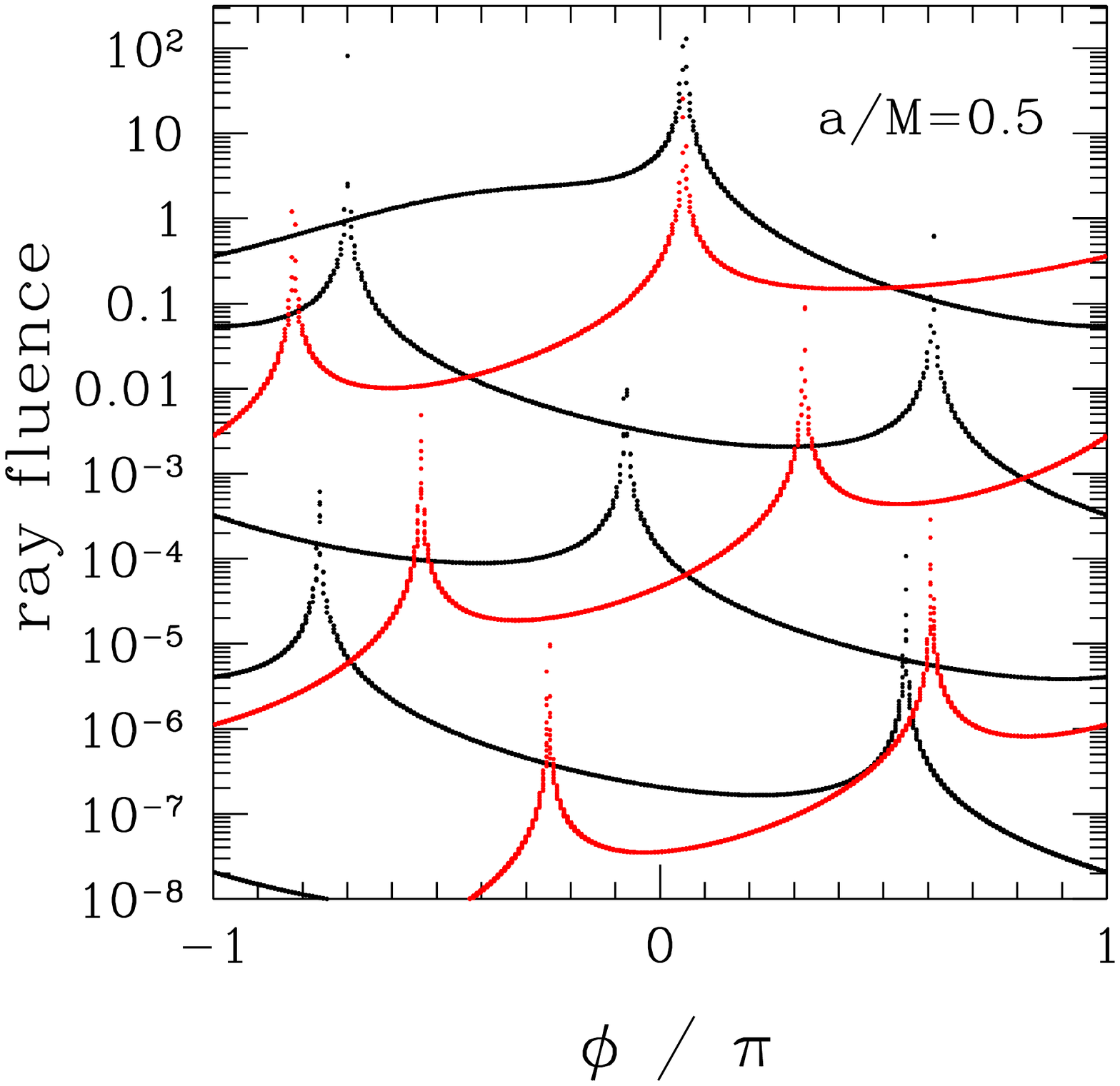}
\vskip -1.2in
\plottwo{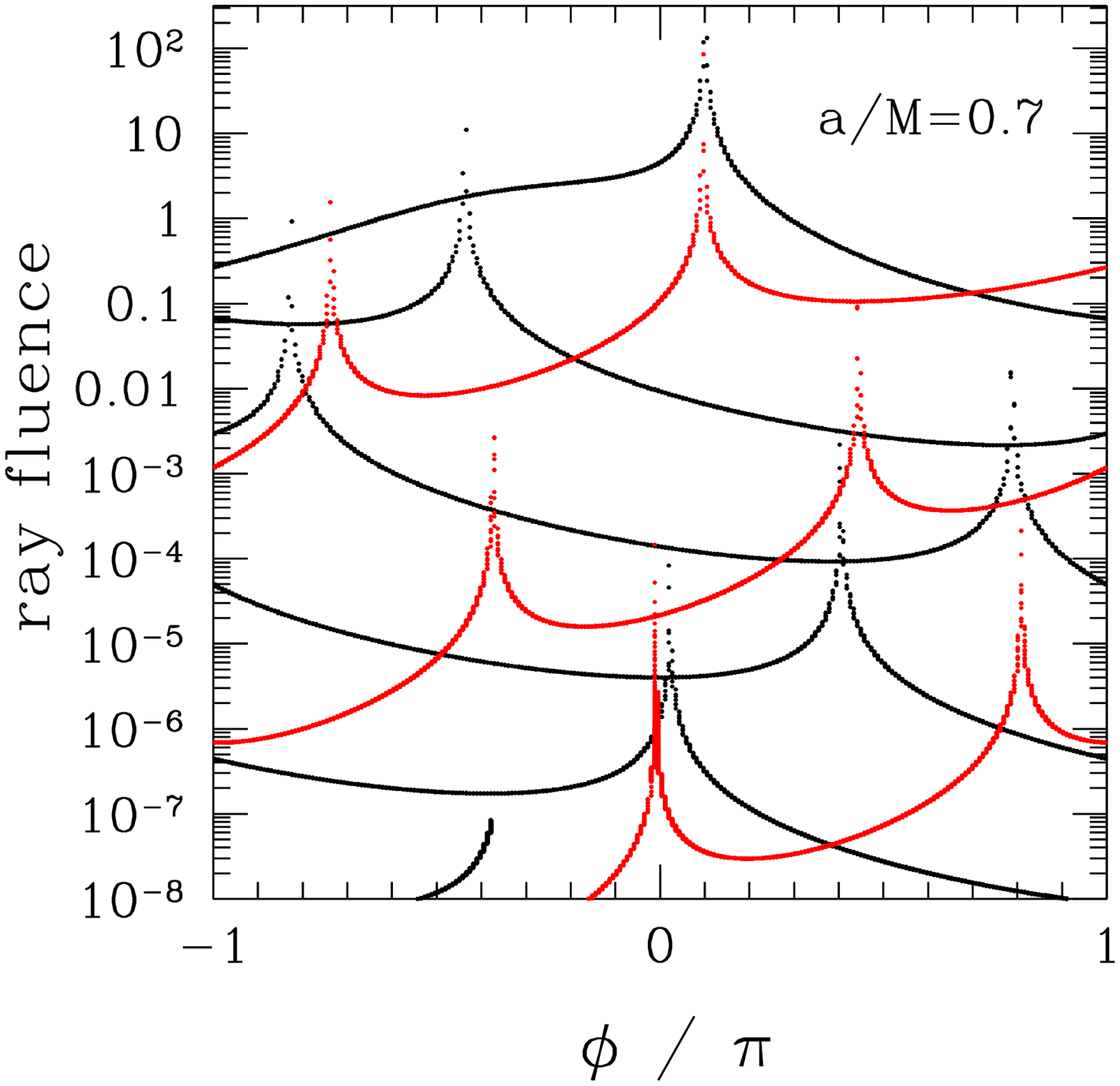}{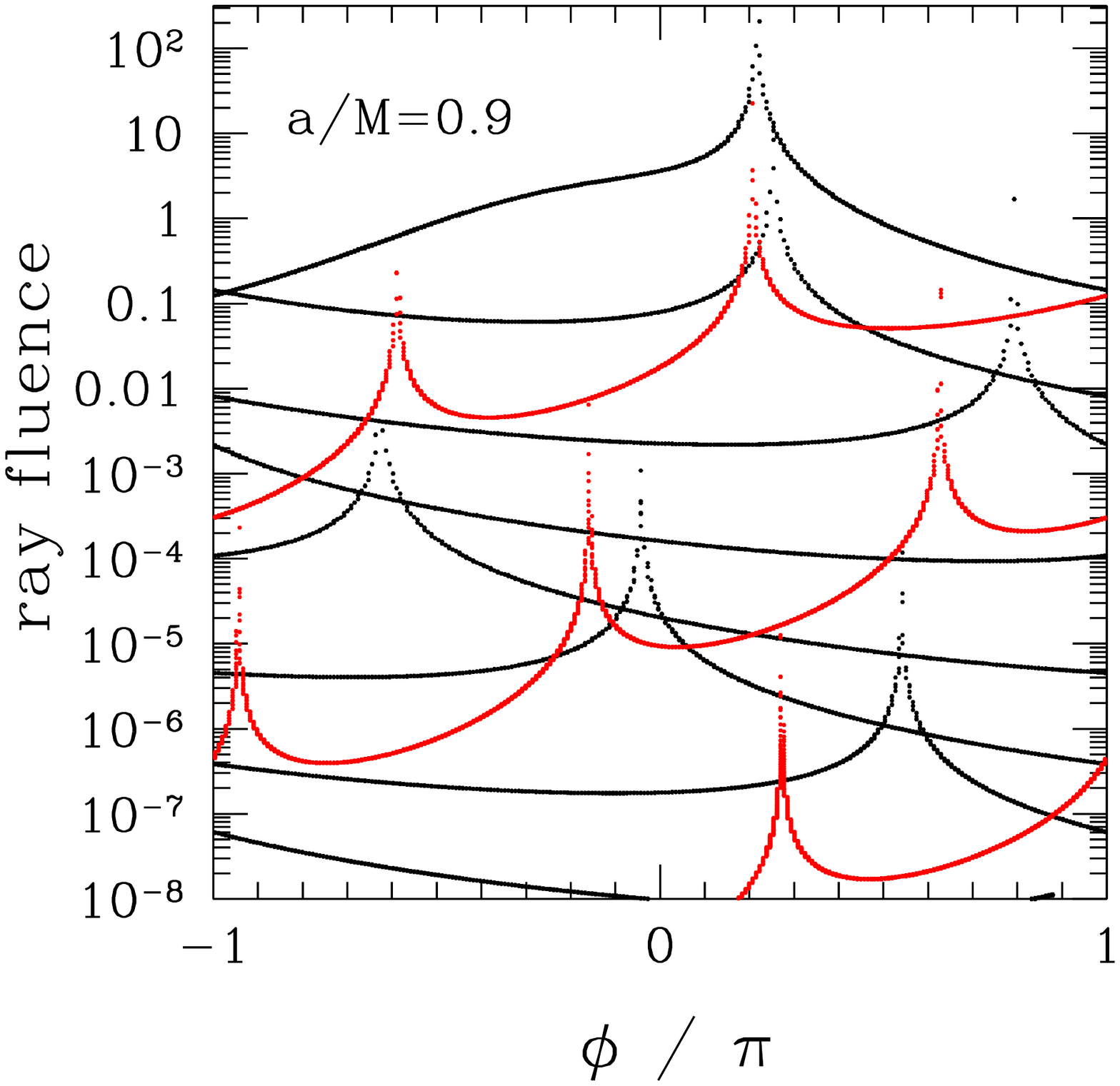}
\vskip -0.6in
\caption{Black and red curves: ray fluence versus azimuth, as measured on the equatorial plane of the BH
at null infinity.  BH spin increases from the top left.  Blue and gold curves:  corresponding energy fluence
after weighting by the energy integral (\ref{eq:integrals2}).  First pulse detected by an observer sitting
at angle $\phi$ corresponds to the top black/blue curve to the left of the top caustic, and the top red/gold curve 
to the right.  Successive pulses of diminishing fluence are emitted at increasing times; see also 
Figure \ref{fig:fluxvst}.  When $a > 0$, the successive caustics shift in angle because of the differing periods
of the prograde and retrograde light rings and the transit times of polar rays.  These curves
are obtained by integrating the Raychaudhuri equations (\ref{eq:Raych2}) for a large number of equatorial rays.
In the cases with higher spins, the contribution from polar wavefronts connecting the caustic points is missing
(see Figure \ref{fig:tvsphi}).}
\vskip .2in
\label{fig:fluxvsphi}
\end{figure}

\begin{figure}
\epsscale{1.}
\plotone{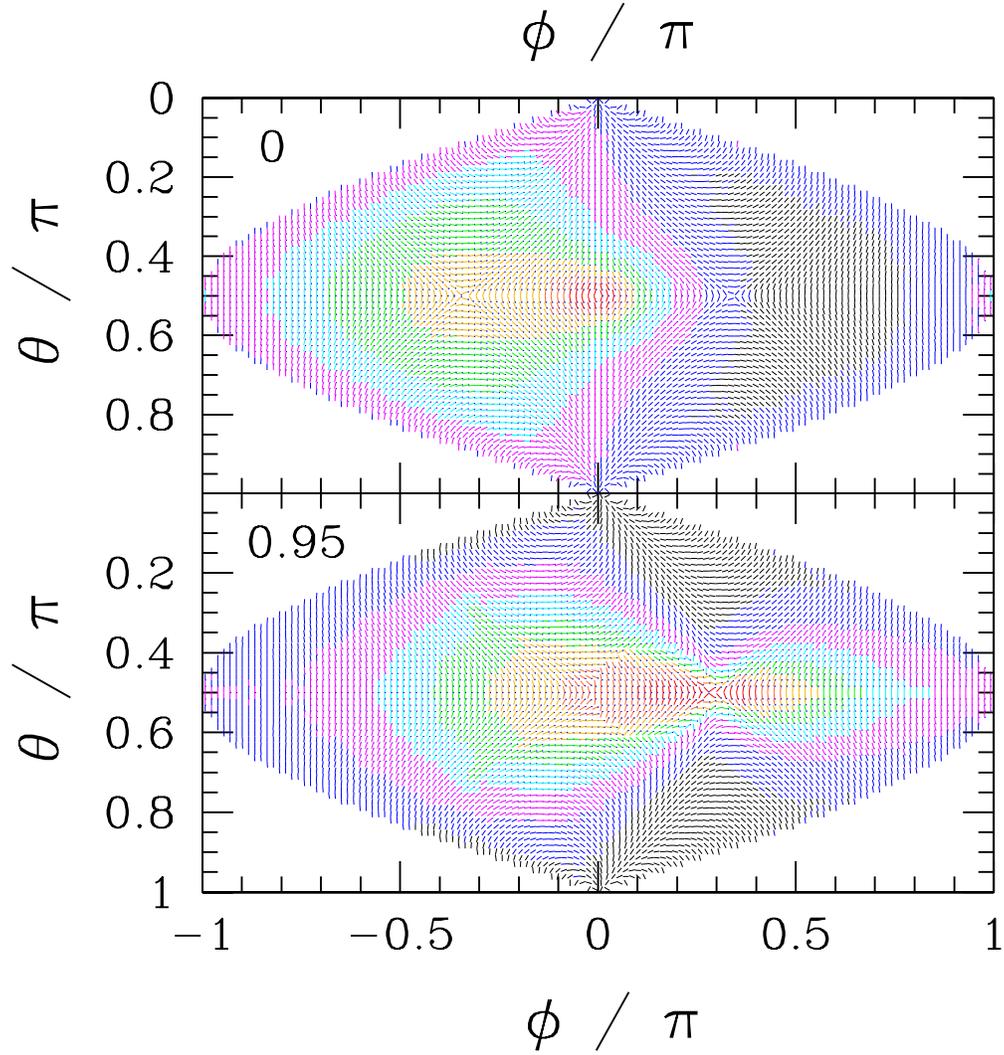}
\vskip -0.9in
\caption{Linear polarization pattern at null infinity, corresponding to emission by the upboosting and reflection of the
ambient toroidal magnetic field near the emission point.  Top panel: $a = 0$;  bottom panel: $a/M = 0.95$.  Color zones are
once again separated by powers of 1.5 (1.7 bottom panel) in intensity.  A relatively uniform polarization 
pattern is seen near the BH equator, and also extending above and below the primary caustic for moderate BH spin.
A differing magnetic configuration at the ISCO (e.g. a poloidal magnetic field) would imply a systematic rotation
of the polarization, but the uniformity in the equatorial polarization pattern would be maintained.  These maps
are obtained by a Monte Carlo procedure, but now with angular resolution 4 times coarser than in Figure \ref{fig:flux_map}.}
\label{fig:polarization_map}
\end{figure}

\begin{figure}
\epsscale{1.1}
\plotone{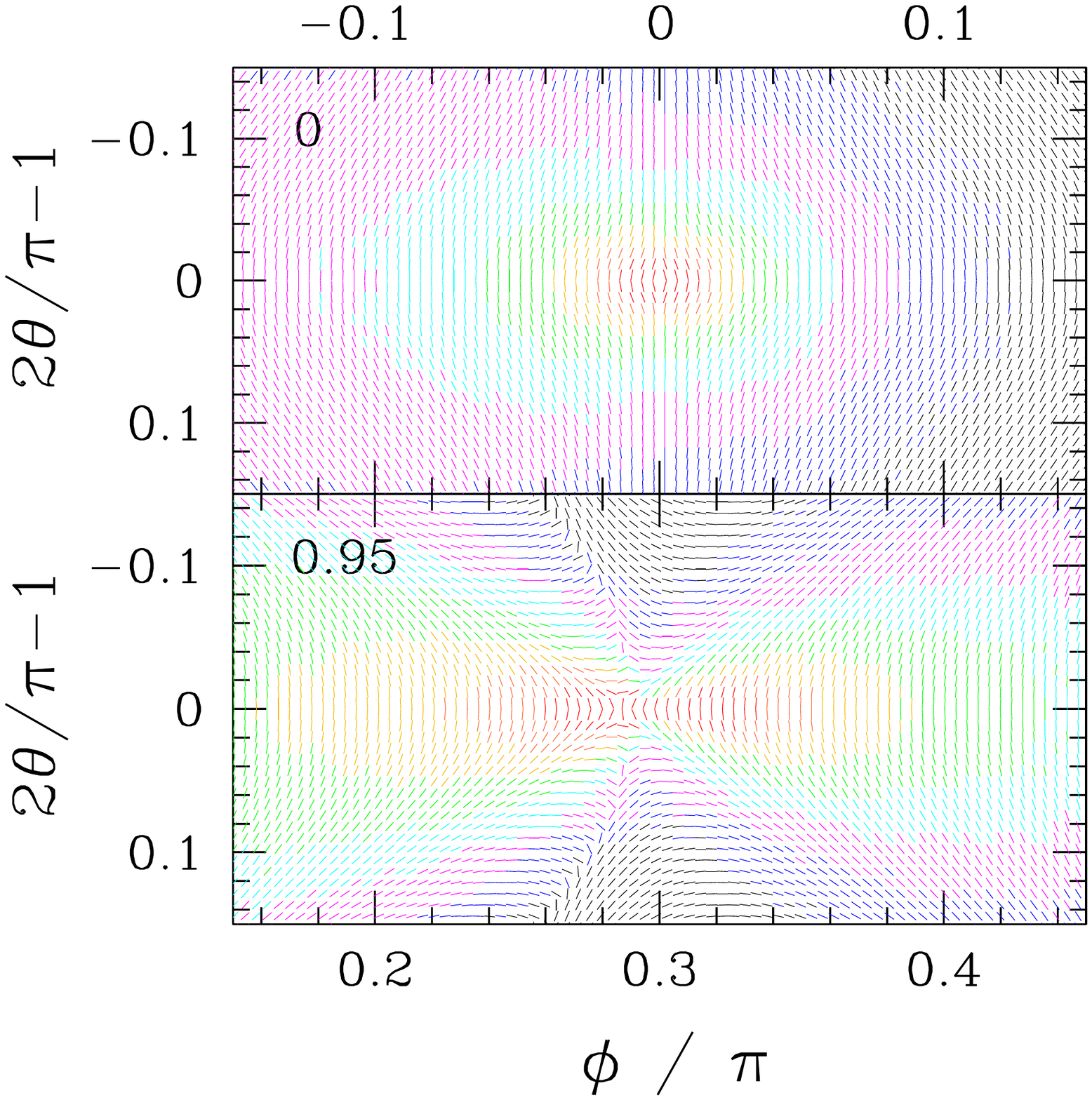}
\vskip -0.9in
\caption{Same as Figure \ref{fig:polarization_map}, but now showing a close up of the polarization 
pattern near the primary caustic, with the same angular resolution as in Figure \ref{fig:flux_map}.}
\vskip .2in
\label{fig:polarization_map2}
\end{figure}

\subsection{Ray Intensity at Null Infinity}\label{s:bright}

The distribution of rays across the sphere at infinity is compared in Figure \ref{fig:flux_map} for various $a/M$.
In a frame other than the emission frame -- in particular the ZAMO frame -- the surface density of rays is inhomogeneous
on a small light sphere surrounding the emission point, being aberrated by a factor 
${\cal D}_{\rm em}^2 = \gamma_{\rm em}^2(1+ \hat k_{\rm em}^{(3)}\beta_{\rm em})^2$.
One may define a ray `fluence' in the case of point emission in time;  at a small affine
distance $\lambda_0$ from the emission point this is
\be
{\cal F}_0 = {\cal D}_{\rm em}^2{{\cal N}\over 4\pi R_{{\rm em},0}^2};  \quad\quad
R_{{\rm em},0} = \lambda_0 (-\widetilde g_{tt})^{1/2} {dt\over d\lambda}.
\ee
In a Monte Carlo calculation, ${\cal N}$ is simply the number of trial rays.  
The ray fluence on a sphere of large radius $r(\lambda) \gg M$ can be normalized to 
emission from a source at rest in flat space,
\be\label{eq:fluence}
{{\cal F}(\lambda)\over {\cal F}_{\rm flat}(\lambda)} = {{\cal D}_{\rm em}^2\over 
\delta A_\perp/\delta A_{\perp,0}} \left[{r(\lambda)\over R_{{\rm em},0}}\right]^2.
\ee
Here, $\delta A_\perp$ is the cross-sectional area of an infinitesimal ray bundle surrounding a given geodesic.  The
Monte Carlo result can be directly tested against an integration of Raychaudhuri's equations; see
Appendix \ref{s:raychcomp}.  

The cumulative Poynting flux transmitted over the small sphere in
the ZAMO frame is related to the total energy ${\cal E}_{\rm em} = k^{(0)}_{\rm em} {\cal N}$ 
released in the emission frame by ${\cal F}_{{\cal E},0} = {\cal D}_{\rm em} k^{(0)}_{\rm em} {\cal F}_0$.   
The Poynting flux transmitted by a ray bundle to a large distance from the BH is obtained by multiplying 
by the energy integral (\ref{eq:integrals}), which is proportional to $k^{(0)}_{\rm em}$ following Equation
(\ref{eq:integrals2}).

In the case of a nonspinning BH, 
gravitational lensing produces a strong intensity peak near the observer azimuth that is antipodal to the emission point,
as well as a secondary peak that is shifted by $180^\circ$.
The ray fluence near the main peak diverges as $\sim |\theta-\pi/2|^{-1}$
for low-to-moderate BH spin (Figure \ref{fig:flux_lat}).  
One also observes in Figure \ref{fig:flux_map} a broad Doppler shift in the intensity contours
in the direction of orbital motion.  

The primary caustic sits precisely at the antipodal point only for a nonspinning BH; differential frame dragging
of prograde and retrograde rays shifts its position when $a > 0$, and the primary and secondary caustics move closer together.
An infinite sequence of caustics is in fact present:  the tertiary caustic aligns with the primary caustic
when $a=0$.   A similar effect was observed by \cite{zg12} in a numerical evolution of point scalar field emission
near a Schwarzschild BH.  Figure \ref{fig:fluxvsphi} demonstrates this effect for nearly equatorial rays.  

Here we also plot the energy fluence at null infinity in the case $a=0$.  
The fluence transported by successive impulses decays in precisely the manner expected for
the lowest radial order spin-1 quasi-normal mode:  the top-left panel of Figure \ref{fig:fluxvsphi}
shows a decay factor equal to $e^{-2\omega_I t_{\rm ring}} = 1/420$, where $\omega_I = 0.1850/2M$ is the imaginary
part of the mode frequency \citep{berti09}, and $t_{\rm ring} = 2\pi\cdot 27^{1/2}\,M$ is the orbital period of the light ring.  
By contrast, when the BH spins, the fluence received by a fixed observer is aperiodically biased by the intervention
of caustic features, which complicates a comparison with the quasi-normal mode.
Further details of the time profile of the delayed pulses are described in Section \ref{s:time}.

The polarization pattern at infinity is shown, at a somewhat lower resolution, in 
Figure \ref{fig:polarization_map}.   One observes that the electric vector maintains
a nearly uniform direction near the rotational equator, excepting for observers oriented close to
the main intensity cusp.  The deflection of the polarization angle near the cusp is shown in more detail
in Figure \ref{fig:polarization_map2}, showing how it strengthens with growing BH spin.

\begin{figure}
\epsscale{1.1}
\plottwo{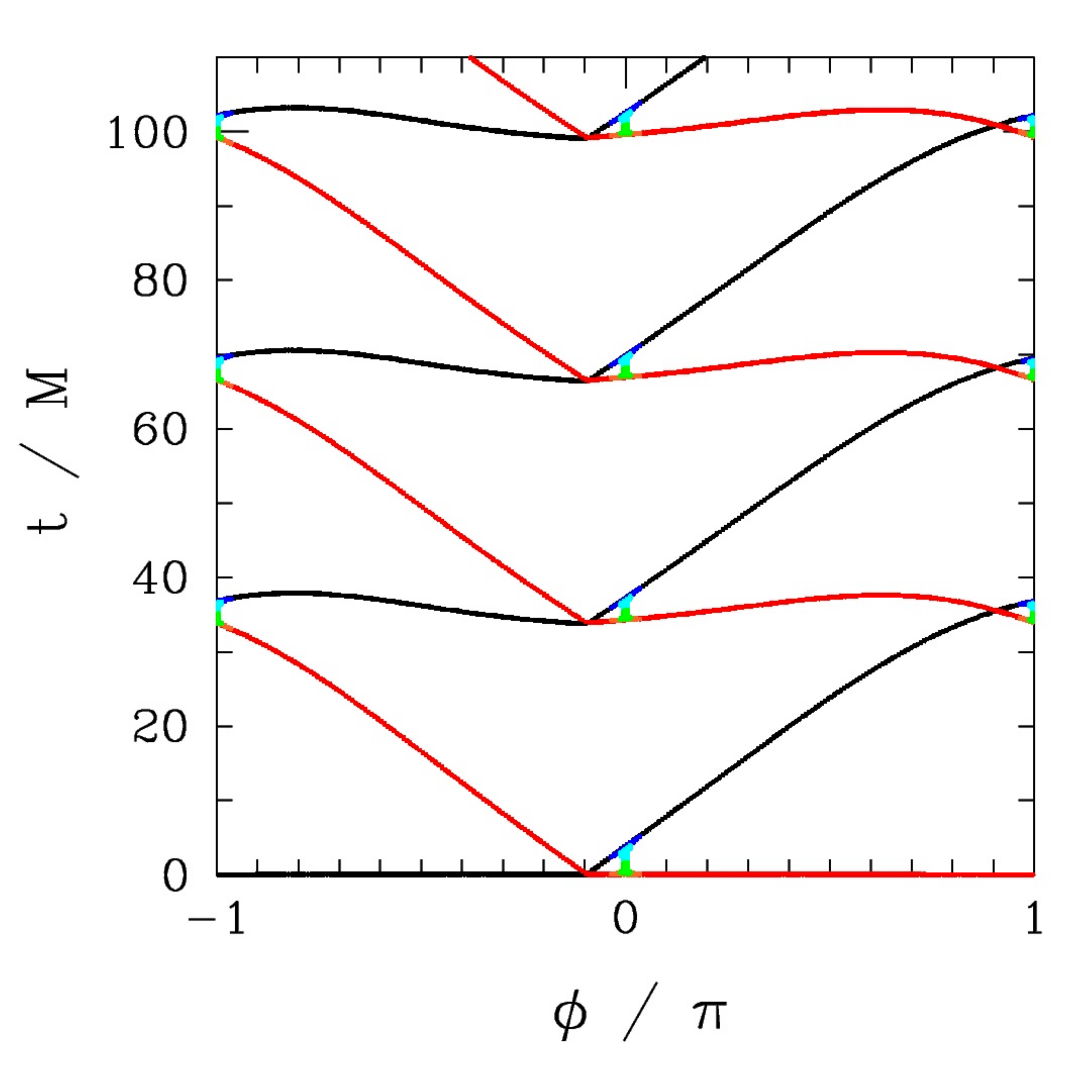}{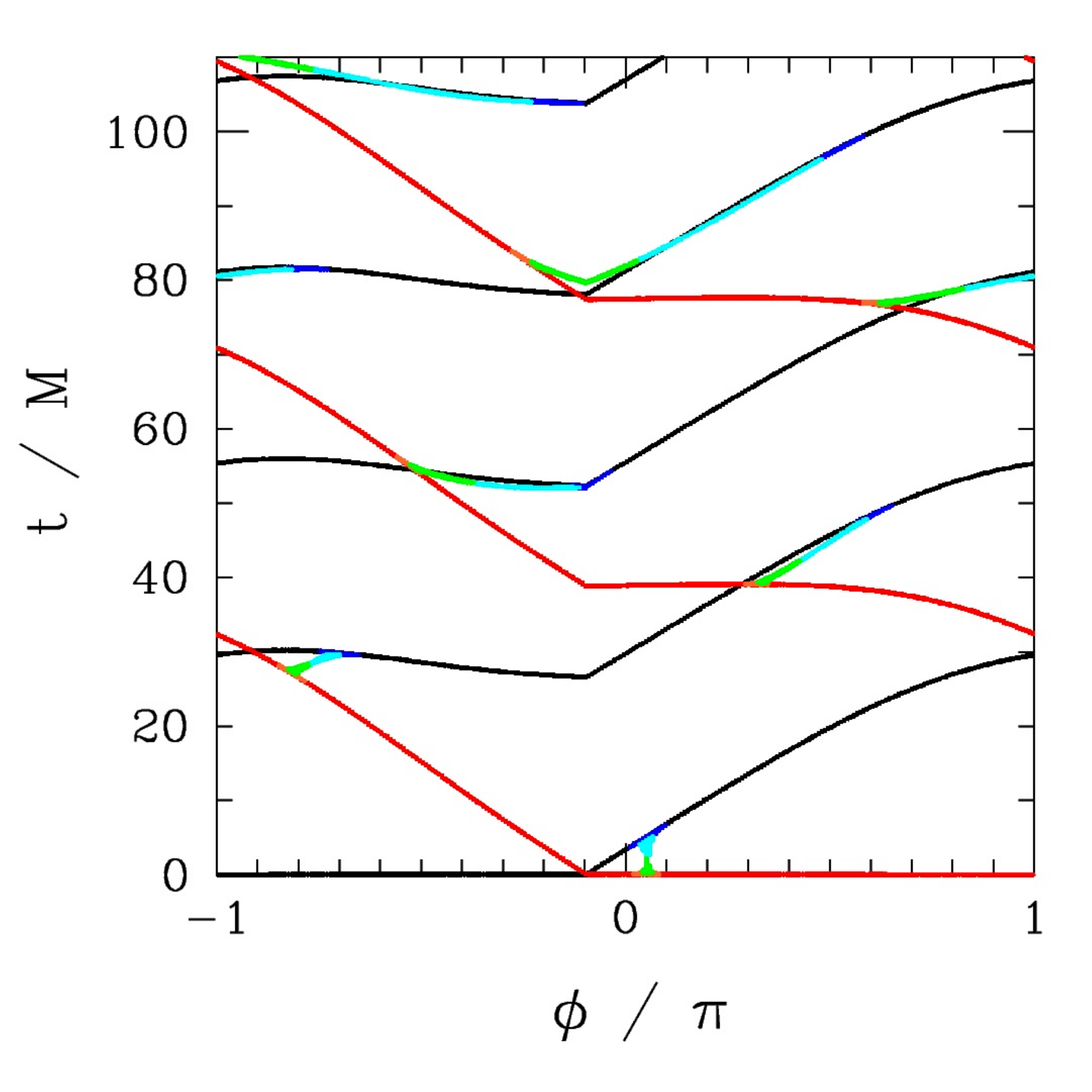}
\plottwo{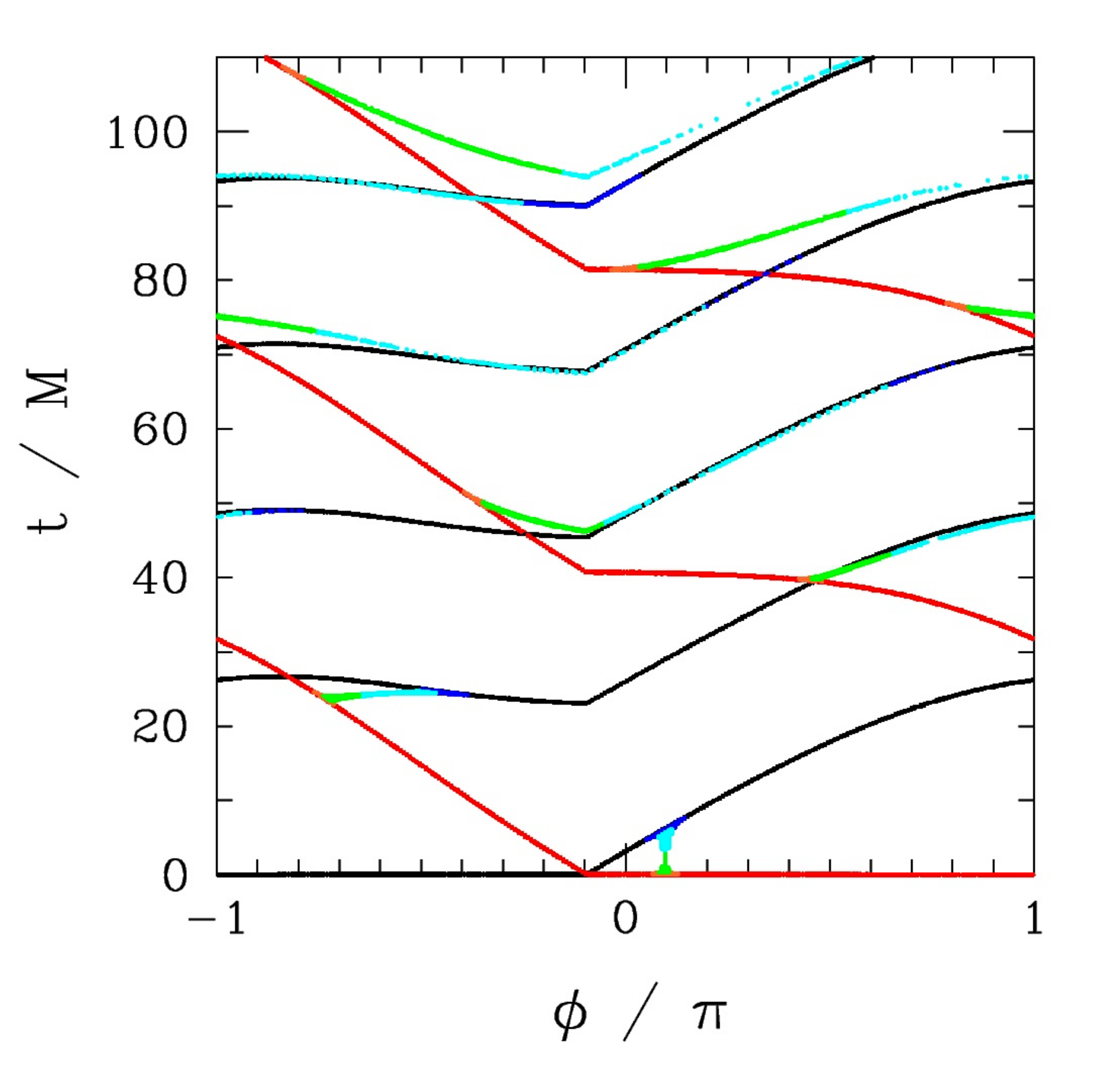}{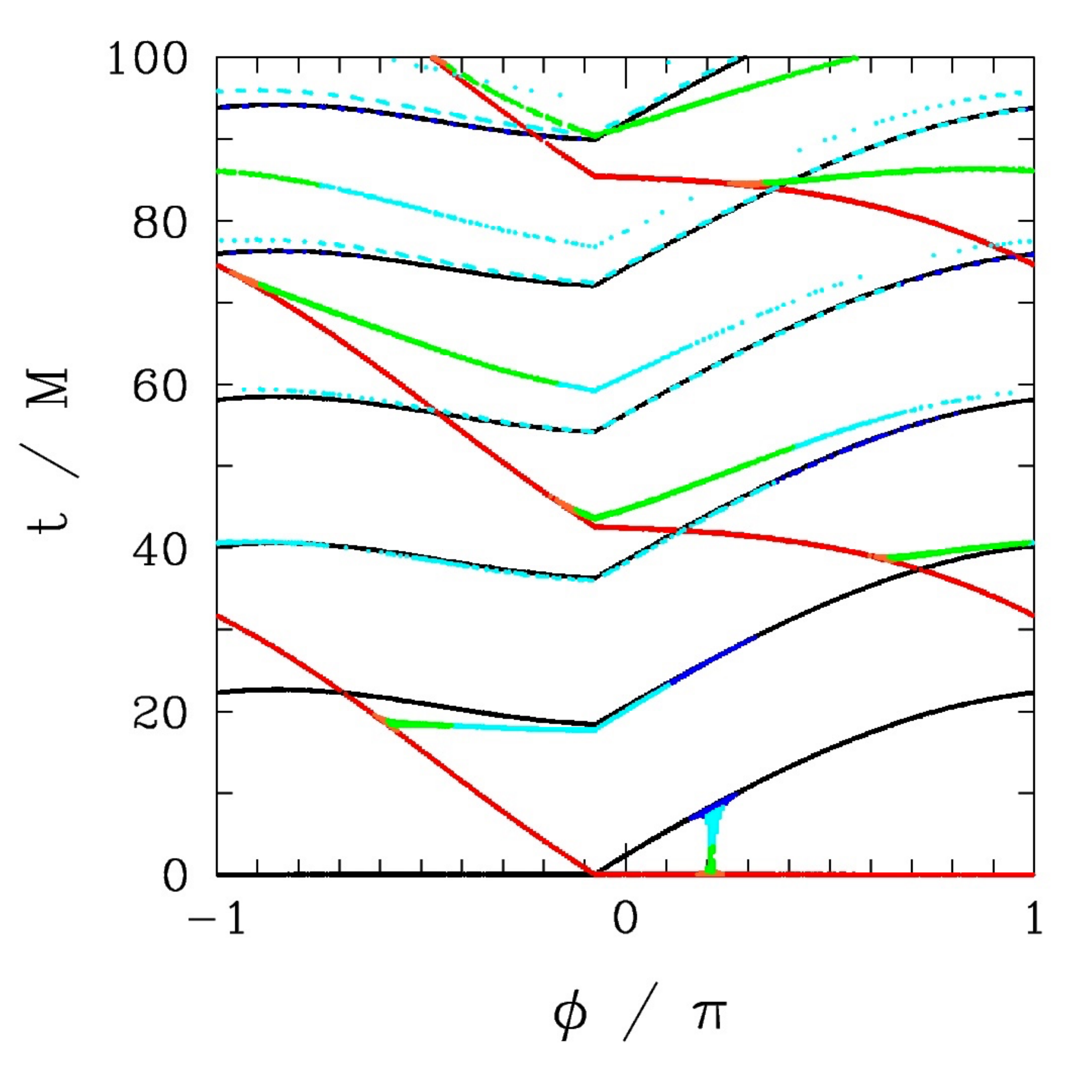}
\caption{Pulse arrival times as a function of the azimuth of an observer positioned
near $\theta = \pi/2$ at a large distance from the BH.  Angle $\phi = 0$ is antipodal
to the emission point on the ISCO.  Top left panel:  $a=0$;  top right panel: $a/M = 0.5$; 
bottom left panel: $a/M = 0.7$;  bottom right panel:  $a/M = 0.9$.  
In all cases the zero of time is shifted to the first pulse detected at a given observing angle.
Black (red) lines denote prograde (retrograde) rays, which remain confined to $|\cos\theta| < 0.1$ from
the emission point all the way to infinity.  The periods of the respective light rings are clearly
visible.  Blue and cyan denote prograde rays with
$0.1 < |\cos\theta|_{\rm max} < 0.5$ and $|\cos\theta|_{\rm max} > 0.5$; orange and green denote the
corresponding retrograde rays.  When $a/M$ is small, most of the delayed pulses detected by an equatorial
observer follow nearly equatorial geodesics, with the exception of caustic features associated with narrow azimuthal 
bundles of rays that reach close to the poles.  The azimuthal range of the polar rays broadens with growing $a/M$;
comparison with Figure \ref{fig:fluxvsphi} shows that the caustics still coincide with the connections of the
cyan/green and black red curves.}
\vskip .2in
\label{fig:tvsphi}
\end{figure}

\begin{figure}
\epsscale{1.15}
\plottwo{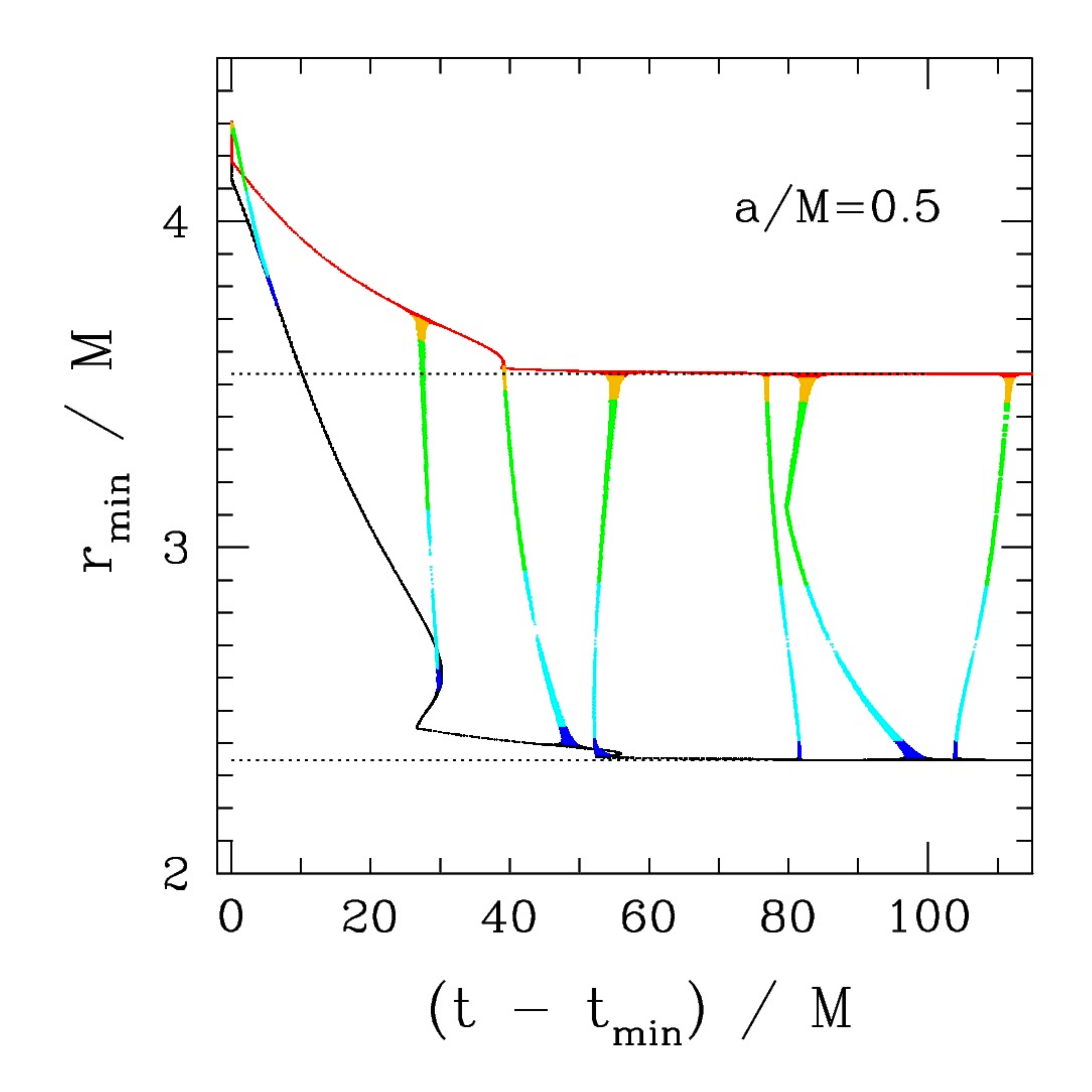}{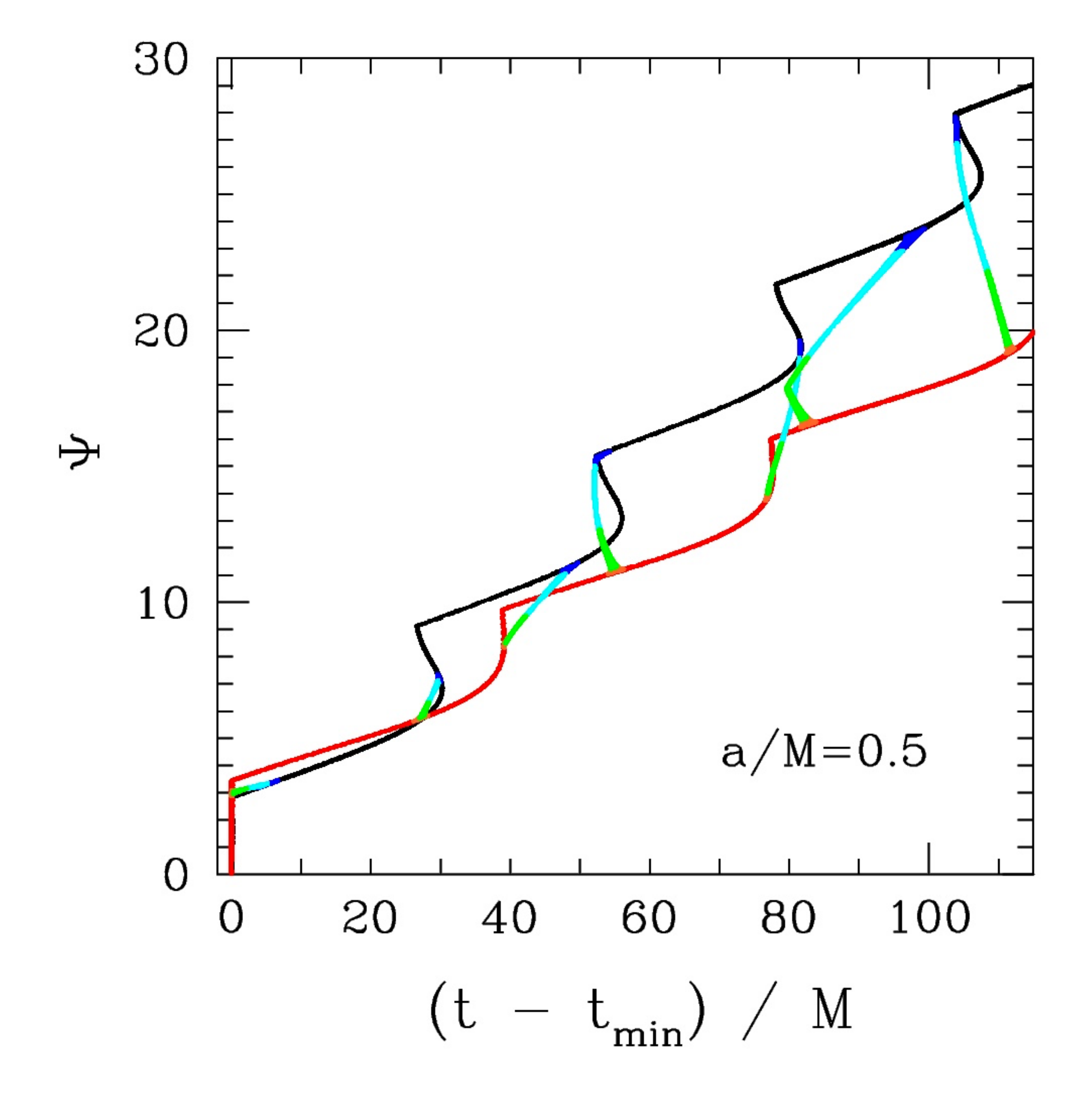}
\caption{Left panel: minimum approach of a ray to the BH versus delay time from the first detected pulse, as 
seen on the BH equator ($\theta = \pi/2$).   Color scheme is the same as in Figure \ref{fig:tvsphi}.  
Black/red points:  prograde/retrograd rays;  blue/cyan/green/gold points denote rays reaching $|\cos\theta| > 0.1$.
Horizontal dotted lines mark the prograde and retrograde photon orbits, here for $a/M = 0.5$.  Right panel:  
winding angle of the ray around the BH.  Rays that closely approach one of the light rings experience a single 
radial bounce ($dr/d\lambda = 0$).}

\vskip .2in
\label{fig:delay_vs_wind}
\end{figure}

\begin{figure}
\epsscale{0.75}
\vskip -0.3in
\plotone{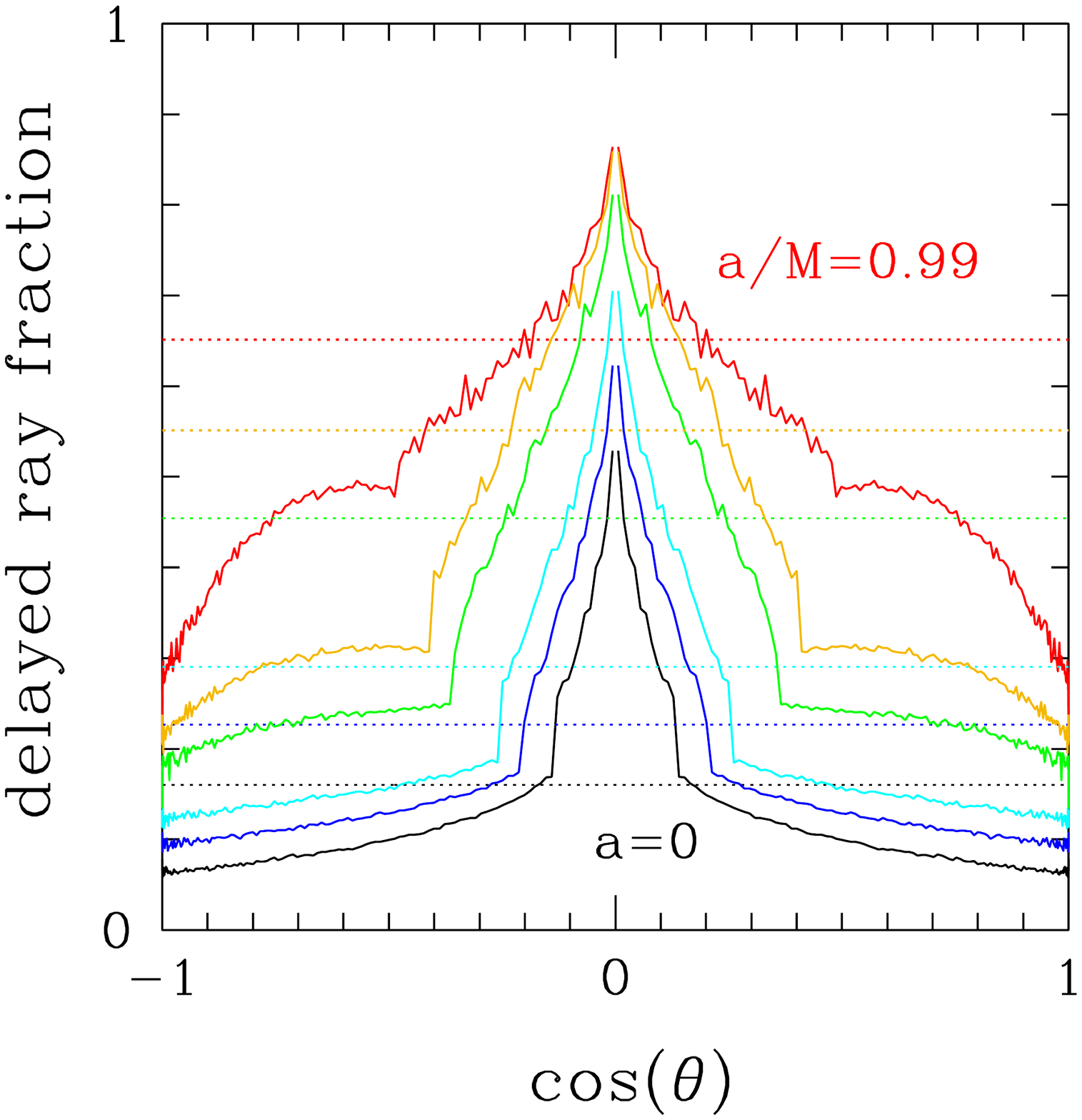}
\vskip -1.1in
\caption{Solid lines: fraction of rays arriving in delayed pulses, as a function of colatitude $\theta$ 
measured on the sphere at infinity, for a range of BH spins ($a/M = 0, 0.5, 0.7, 0.9, 0.95, 0.99$, bottom to top).
Dotted horizontal lines:  fraction of rays arriving in delayed pulses, averaged over colatitude.
The ray fluence is weighted more strongly to the equator for higher $a/M$, hence each dotted line is not
a direct $\theta$-average of the corresponding solid line.}
\label{fig:tailfraction}
\end{figure}

\subsection{Pattern of Delayed Pulses:  Equatorial vs. Polar Rays}\label{s:time}

The zero of time is adjusted, independently for each observer direction, to the arrival of the initial
impulse.  The sequence of arriving pulses seen by observers aligned with the BH equator, at random values of $\phi$
relative to the emission point, is shown in Figure \ref{fig:tvsphi}.  The initial pulse connects at a particular
azimuth with delayed equatorial rays (black and red curves) and at a slightly different azimuth with a 
narrow bundle of polar rays (green and cyan curves).  This confluence of equatorial and polar rays 
coincides with the first, brightest caustic (Figure \ref{fig:flux_map}).  The same phenomenon is repeated
for the delayed pulses.  However, as the BH spin grows, the rays reaching away from the equator
(the green and cyan curves) grow in azimuthal extent:  they connect separate caustic features on the prograde and
retrograde rays that are offset from each other in azimuth (compare Figure \ref{fig:fluxvsphi}).

Except for a nonspinning BH, the prograde (black) and retrograde (red) delayed pulses show different
periods, which match the orbital periods of prograde and retrograde equatorial photon orbits (e.g. $2\pi\cdot 27^{1/2}\,M
\simeq 32.6\,M$ in the Schwarzschild case).
This is confirmed by plotting the relation between delay time and winding angle or the minimum approach of 
the ray to the BH (Figure \ref{fig:delay_vs_wind}).  Rays that closely approach one of the light rings 
experience a single radial bounce ($dr/d\lambda = 0$).

For larger BH spins, a larger fraction of rays also arrive in delayed pulses, especially those rays propagating
near the BH equator (Figure \ref{fig:tailfraction}).

\begin{figure}
\epsscale{0.95}
\plottwo{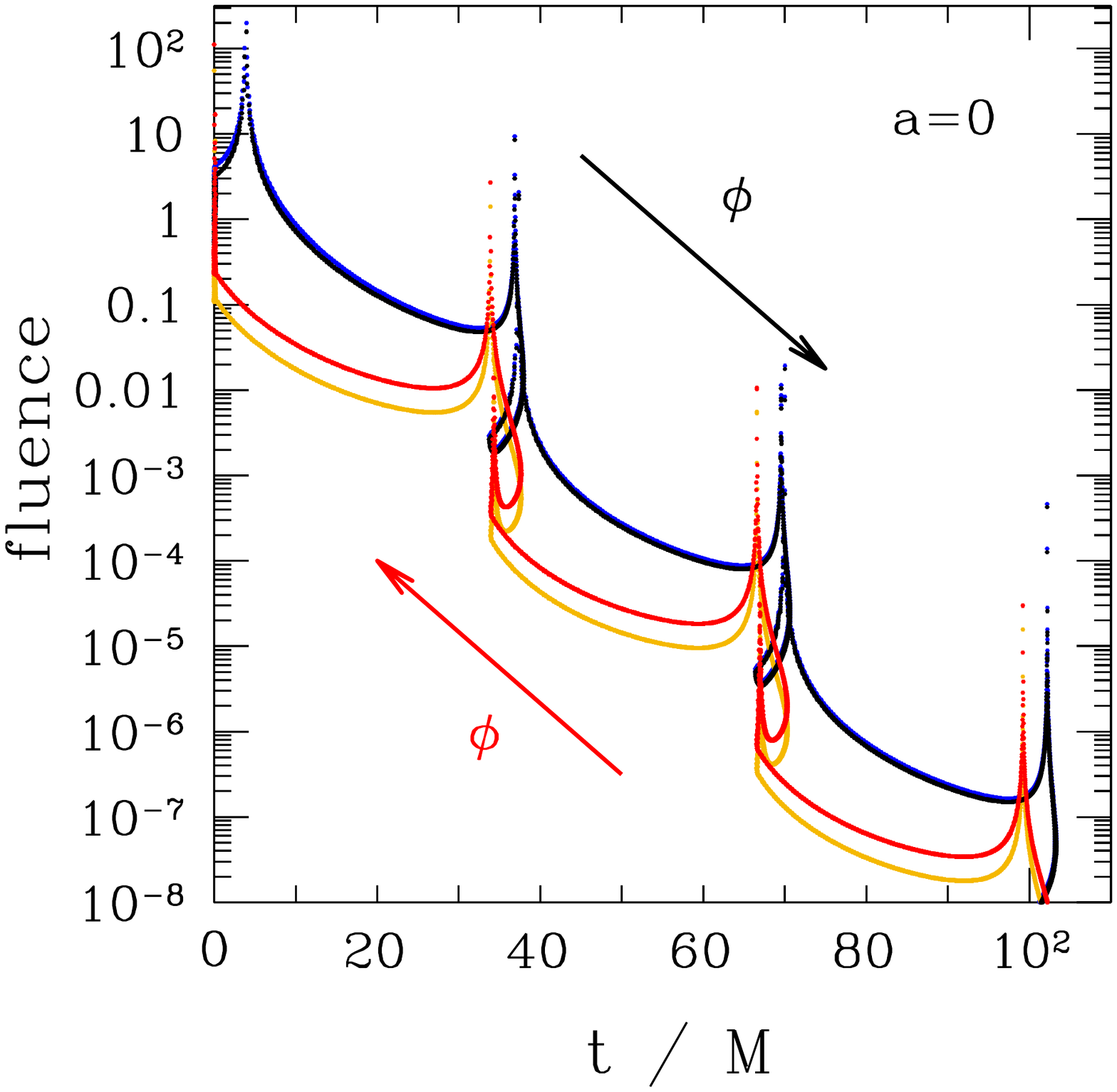}{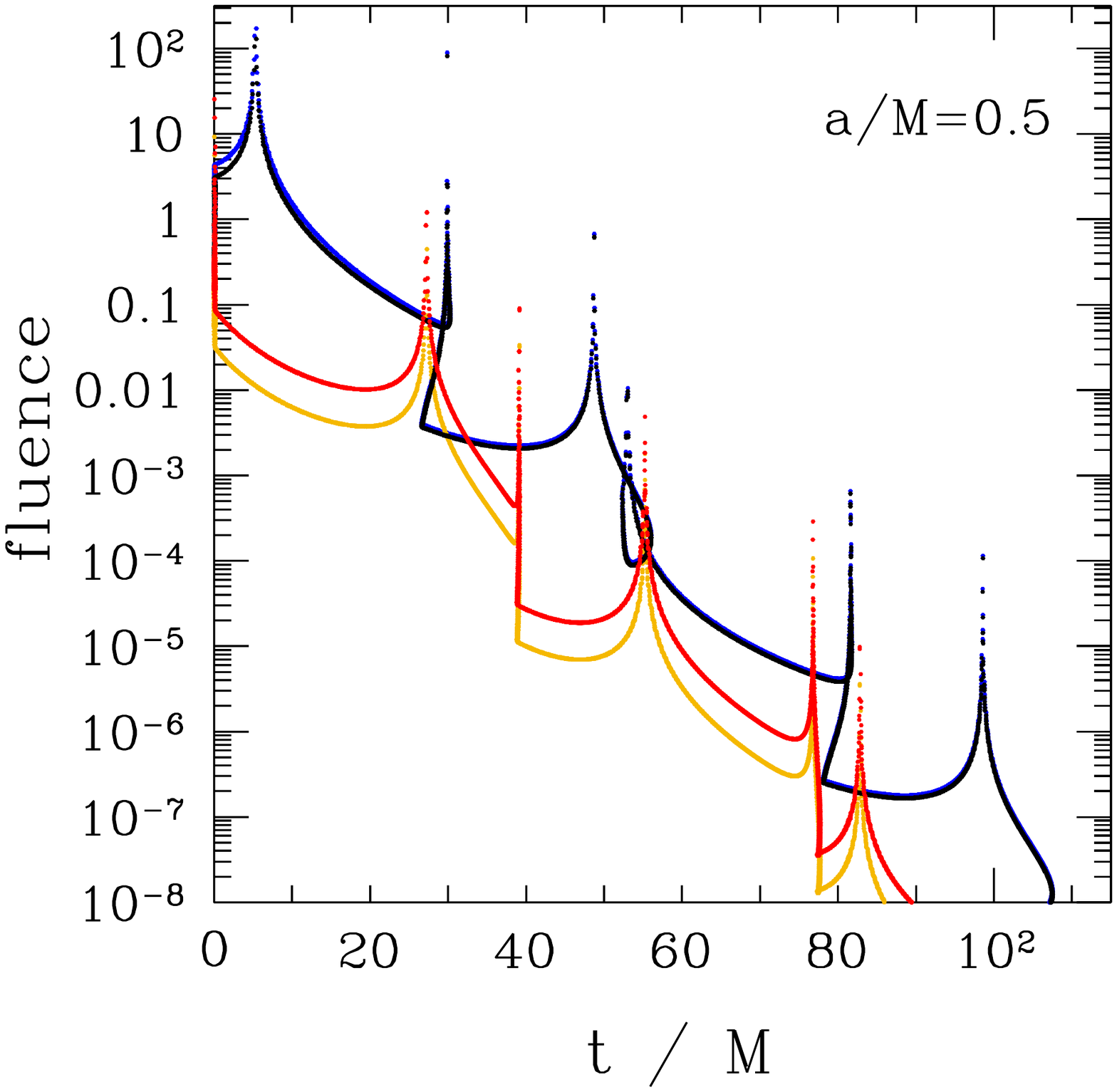}
\vskip -1.2in
\plottwo{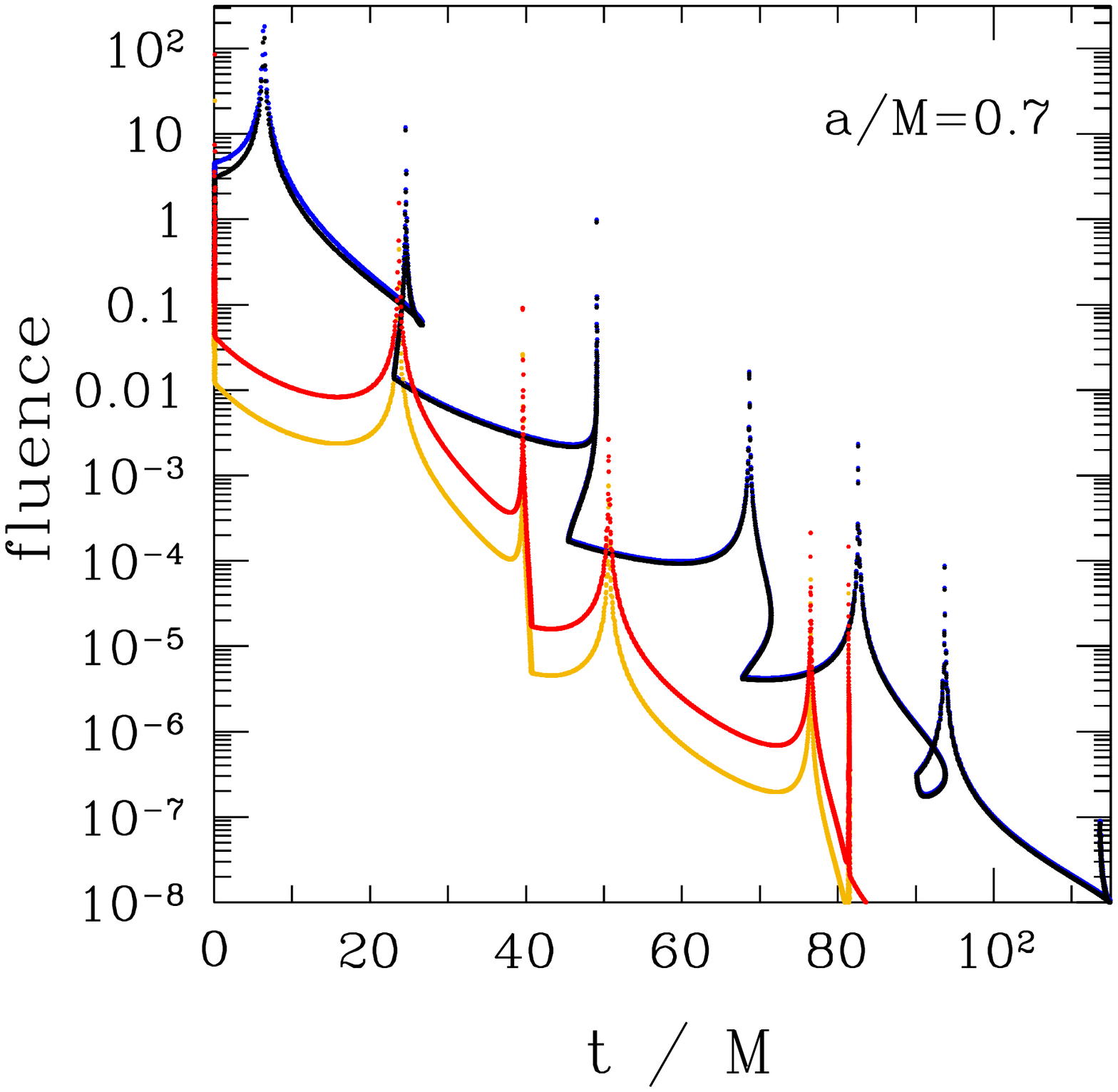}{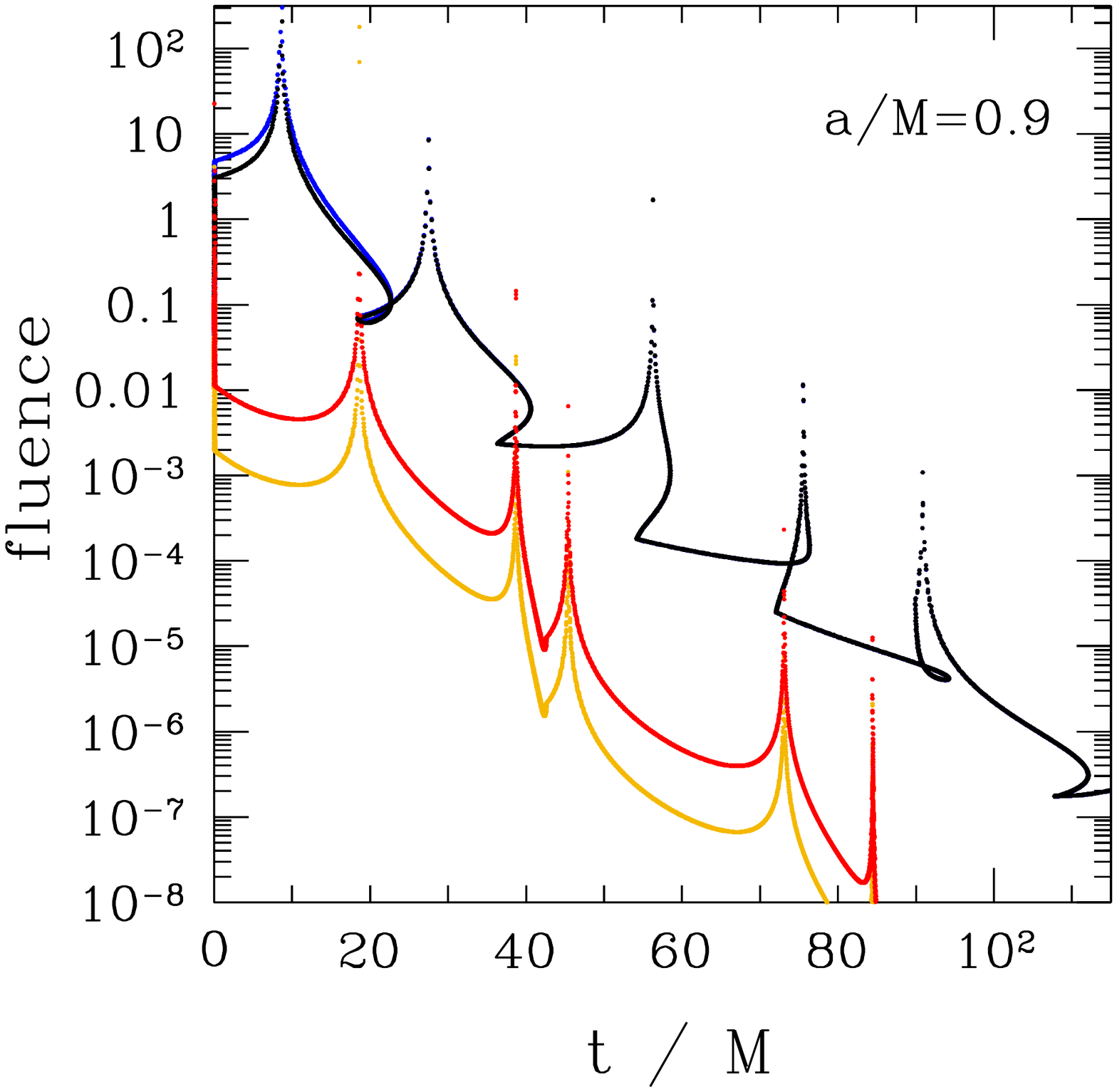}
\vskip -1.2in
\plottwo{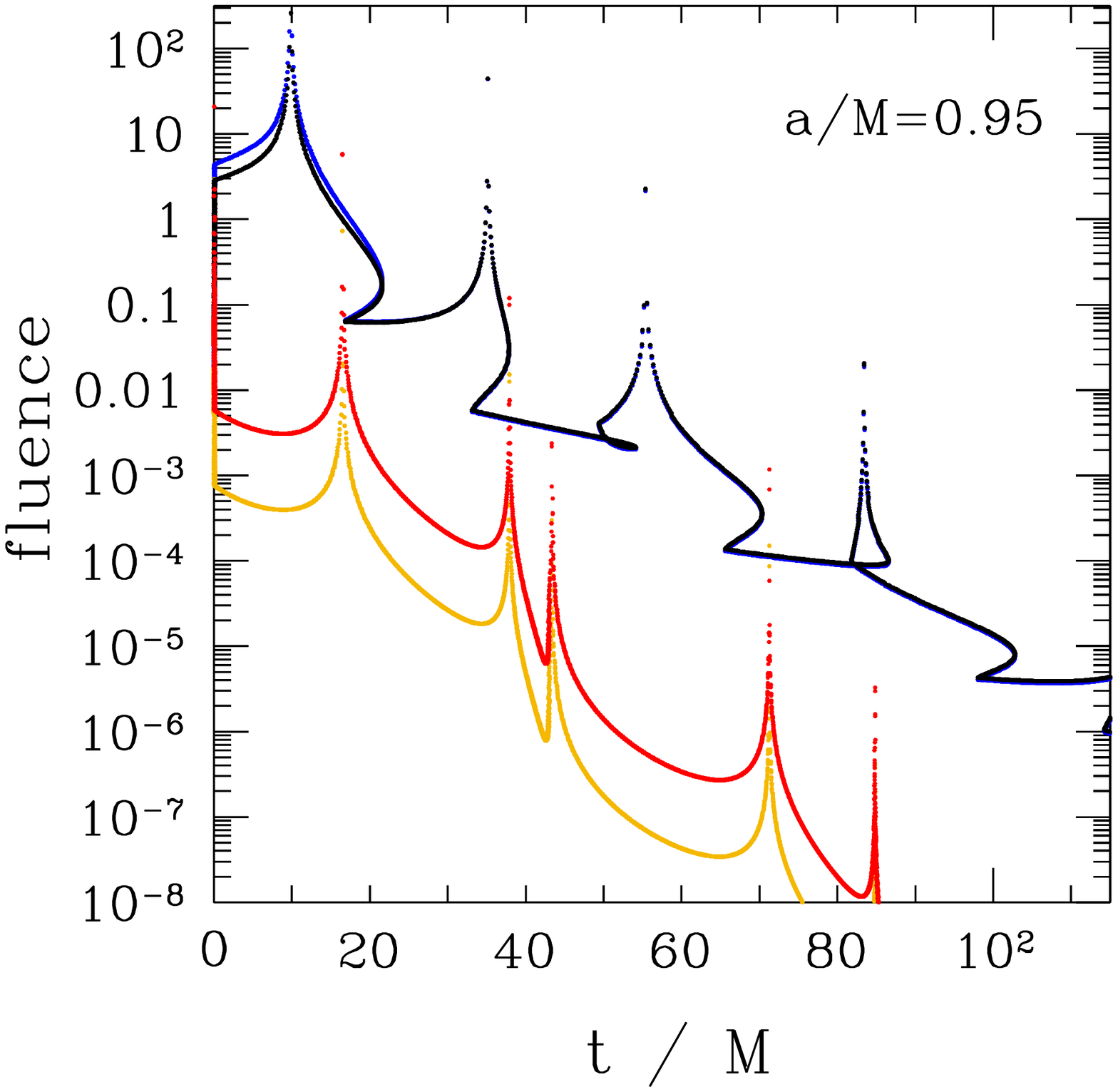}{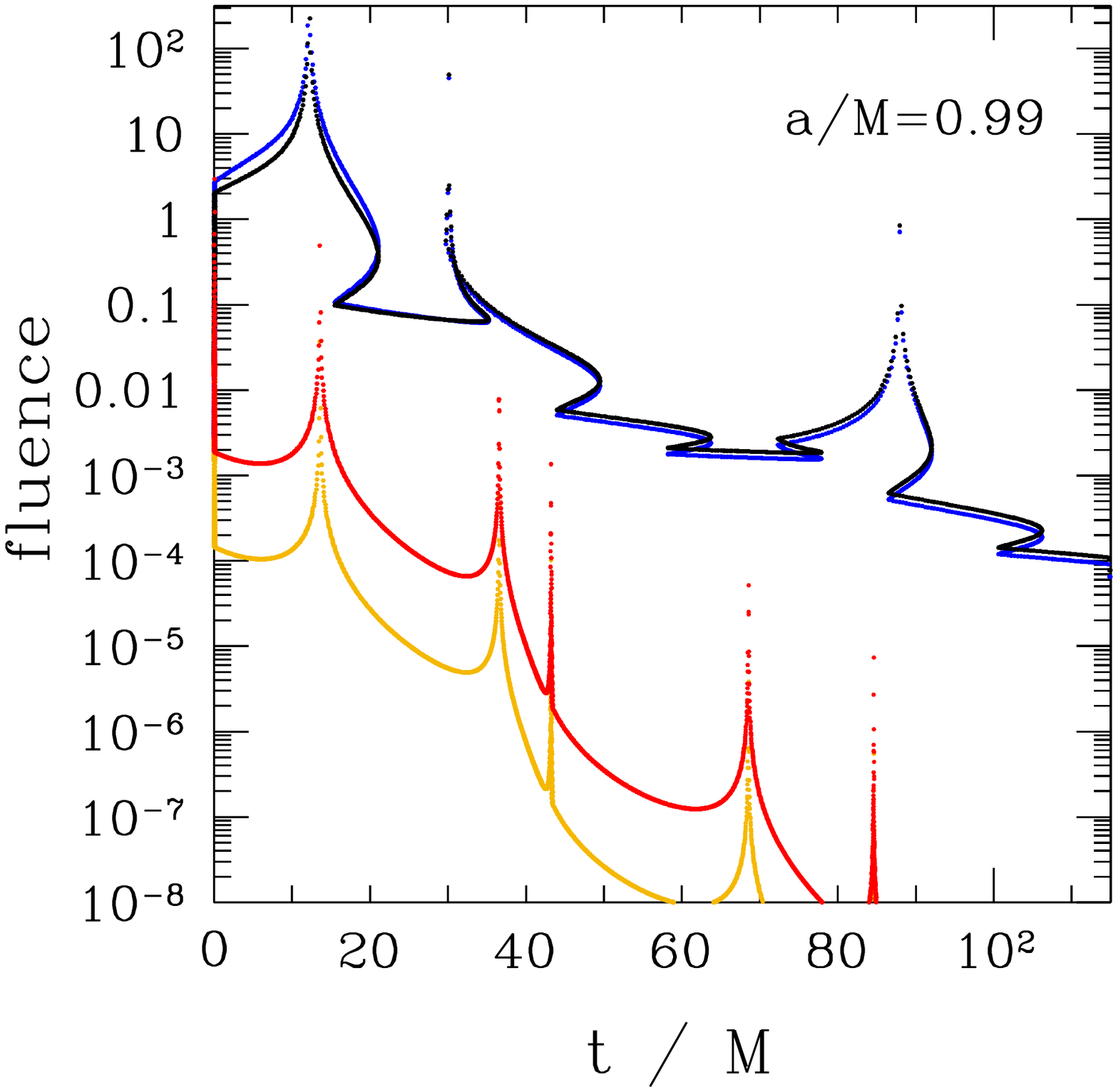}
\vskip -0.5in
\caption{Same as Figure \ref{fig:tvsphi}, but now fluence vs. time, for observers positioned at colatitude
$\theta = \pi/2$.  Black/red points show the prograde/retrograde ray fluence and track the corresponding curves on the 
$t$-$\phi$ plot, with angles $\phi = -\pi$ and $\pi$ identified.  Blue and gold points show the energy fluence, 
weighted by the energy integral of each ray.}
\label{fig:fluxvst}
\vskip .5in
\end{figure}

\section{Defocusing of Equatorial Rays}\label{s:raych}

Equatorial rays offer more extensive and detailed information about the delayed light pulses produced by a point
explosion than can be obtained with a brute force Monte Carlo procedure.  Near the BH equator, it is simple to follow the ray
expansion through turning points near one of the light rings.  A comparison between the results of the two methods
also provides a sharp test of both (Appendix \ref{s:raychcomp}).

The Raychaudhuri equations simplify for equatorial null geodesics.  We work with a tetrad $e^\mu_A$ that projects
four-dimensional coordinates $x^\mu$ onto a transverse $2\times 2$ dimensional space labeling the physical
polarization degrees of freedom.  The ray shear tensor $B_{AB} = e^\mu_A e^\nu_B \nabla_\mu k_\nu$ is decomposed
into an isotropic expansion $\theta = d\ln \delta A_\perp/d\lambda = \delta^{AB}B_{AB}$ (where
$\delta A_\perp$ is the cross section of a narrow ray bundle) and a trace-free component 
$\sigma_{AB} = {1\over 2}(B_{AB} + B_{BA} - \theta\delta_{AB})$.  Near the emission point on the ISCO,
the off-diagonal components of $B_{AB}$ vanish, and so the rotation can be set to zero along the ray.
In the case of an equatorial geodesic, the off-diagonal components of $\sigma_{AB}$ vanish exactly by symmetry.
Then the (upper) diagonal component $\sigma_+$ evolves along with the expansion as \citep{poisson04}
\ba\label{eq:Raych}
{d\theta\over d\lambda} &=& -{1\over 2}\theta^2 - 2\sigma_+^2;\nn
{d\sigma_+\over d\lambda} &=& - \theta\sigma_+ - R_{\alpha\mu\beta\nu}k^\alpha e^\mu_1 k^\beta e^\nu_1
\equiv -\theta\sigma_+ - C_+.
\ea
Here the polarization tetrad must be parallel propagated along the ray,
$de^\mu_A/d\lambda = -\Gamma^\mu_{\alpha\beta} e^\alpha_A dx^\beta/d\lambda$.   A small affine distance
$\lambda_0$ from the emission point, the expansion is initialized as $\theta = 2/\lambda_0$.  The
polarization is initialized in the emission frame as 
\ba
e^{(i)}_1 &=& -(\hat k_{\rm em}\times\hat\theta)^{(i)} /|\hat k_{\rm em}\times \hat\theta|; \nn
e^{(i)}_2 &=& (\hat k_{\rm em} \times \vec e_1)^{(i)},
\ea
where $\hat k^{(i)}_{\rm em} \equiv k_{\rm em}^{(i)}/k^{(0)}_{\rm em}$, and then transformed into the B-L frame
using Equation (\ref{eq:tet}), $e^\mu_A = e^\mu_a e^a_A$.  The equations (\ref{eq:Raych}) may be combined to give
\be\label{eq:Raych2}
{d^2y_\pm\over d\lambda^2} = \mp C_+ y_\pm;  \quad\quad y_\pm(\lambda,\lambda_0) 
\equiv \exp\left[{1\over 2}\int_{\lambda_0}^\lambda (\theta \pm 2\sigma_+) d\lambda'\right].
\ee
After integrating these equations along an equatorial ray, one obtains the expansion
\be
{\delta A_\perp(\lambda)\over \delta A_\perp(\lambda_0)} = {\rm abs}(y_+ y_-).
\ee
The ray fluence at null infinity may then be compared with the result for emission by a stationary source in flat
space using Equation (\ref{eq:fluence}).

The ray fluence is plotted for a large number of randomly chosen equatorial rays in Figure \ref{fig:fluxvsphi} 
(as a function of the relative azimuth of the observer) and in Figure \ref{fig:fluxvst} (as a function of 
the arrival time of the pulse).  One sees that light pulses delayed by successive circuits of the BH form 
a caustic at the same value of $\phi$ only in the case of vanishing BH spin.   Otherwise, the prograde and retrograde 
light-ring periods differ from each other and from the transit times of polar rays, so that the successive caustics
shift in phase.

\begin{figure}
\vskip -0.3in
\epsscale{1.16}
\plottwo{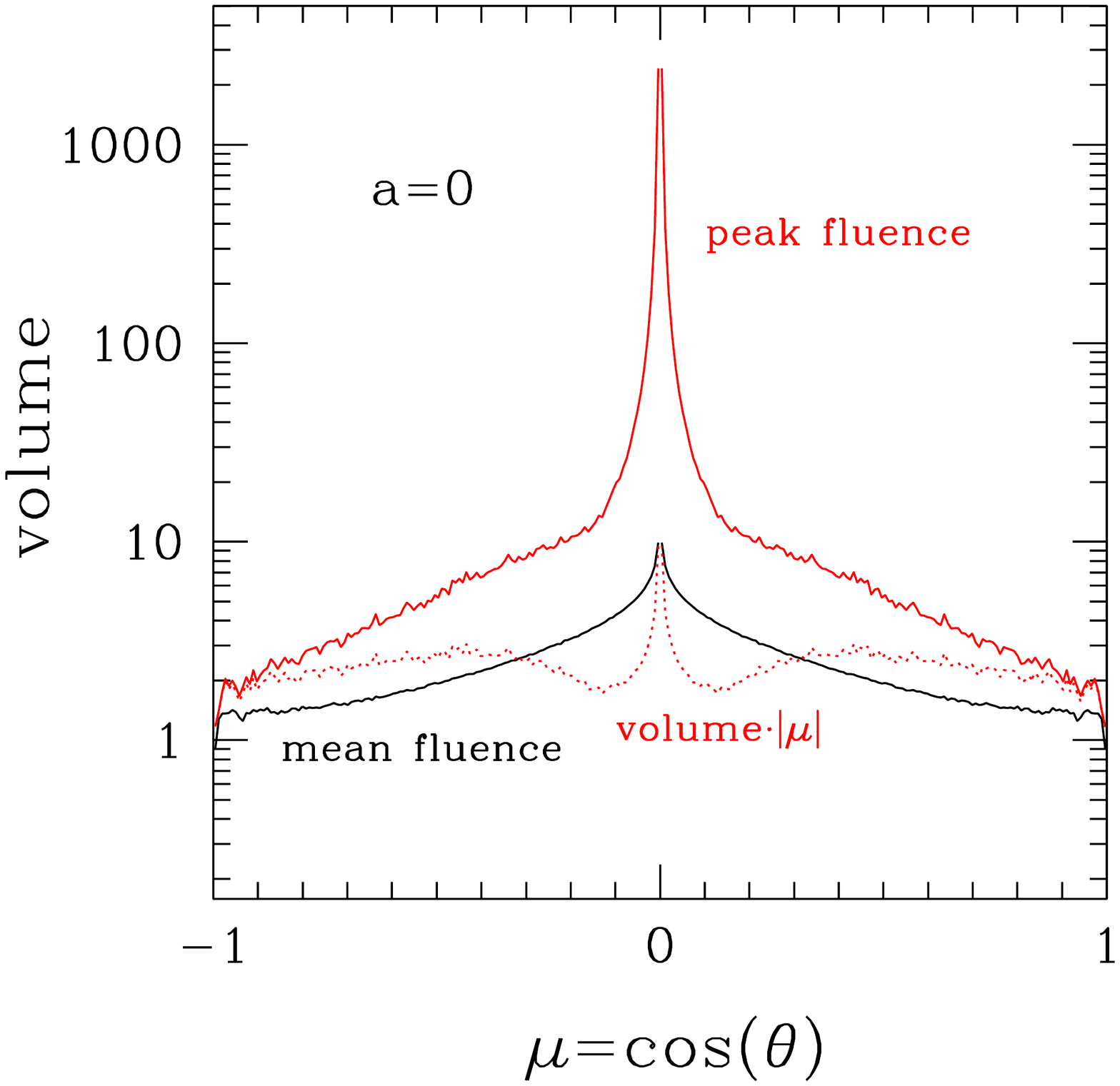}{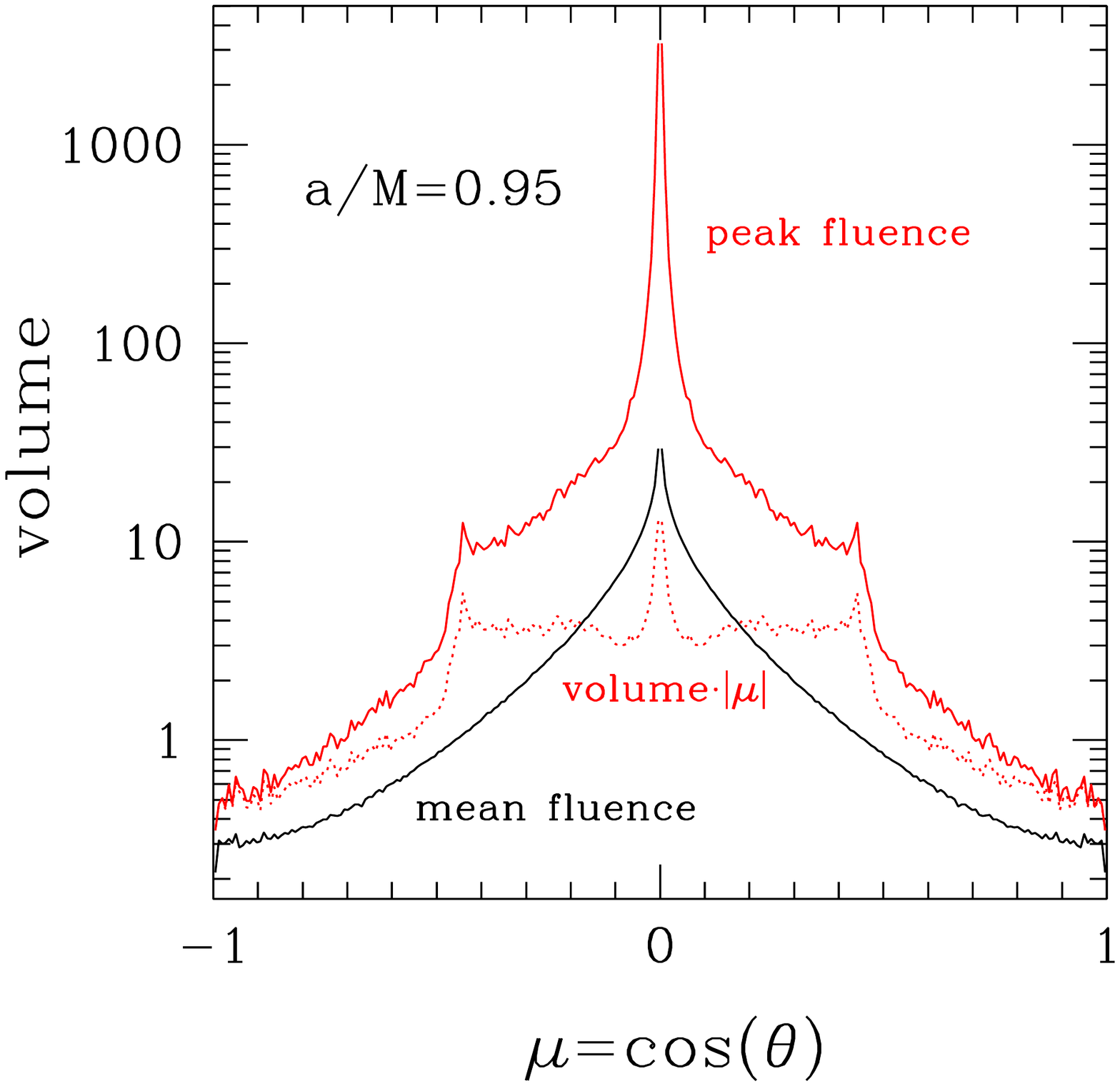}
\vskip -0.8in
\caption{Normalized euclidean volume $V$ corresponding to a fixed threshold pulse fluence, as detected
by an observer at fixed BH latitude.  Black curve: fluence is averaged over successive pulses emitted at random 
orbital phase with respect to the observer.   Red curve:  volume corresponding to the peak fluence
detected at a given latitude.  The cutoff in volume (and fluence) at $\mu=0$ is an artifact of the finite size of the angular
grid used to map outgoing rays.  Dotted curve:  differential volume $V \cdot \mu$.  Left panel:  
$a = 0$;  right panel: $a/M = 0.95$.}
\vskip .2in
\label{fig:volume}
\end{figure}

\section{Lensing Bias:  Detection Rate vs. BH Orientation}\label{s:bias}

Consider now the possibility that repeated electromagnetic point explosions occur close to the ISCO of a BH.
This is the case in the model of FRBs described in \cite{thompson17a,thompson17b}.  Here an observer aligned with
the BH equator ($\theta = \pi/2$) will see on occasion a very bright pulse, when the emission point
lies diagonally opposite to the observer's direction.  If the explosion energy is limited to some maximum value
(as it is in the model just described, as determined by the mass of annihilating dark matter particles),
then a radio telescope will be sensitive to emission from a much greater volume when the BH spin is aligned 
with the plane of the sky, as compared with emission from a BH with a more randomly directed spin.
Figure \ref{fig:volume} shows that the detection volume in Euclidean space has a strong spike at $\theta = \pi/2$.

\begin{figure}
\vskip -0.3in
\epsscale{0.7}
\plotone{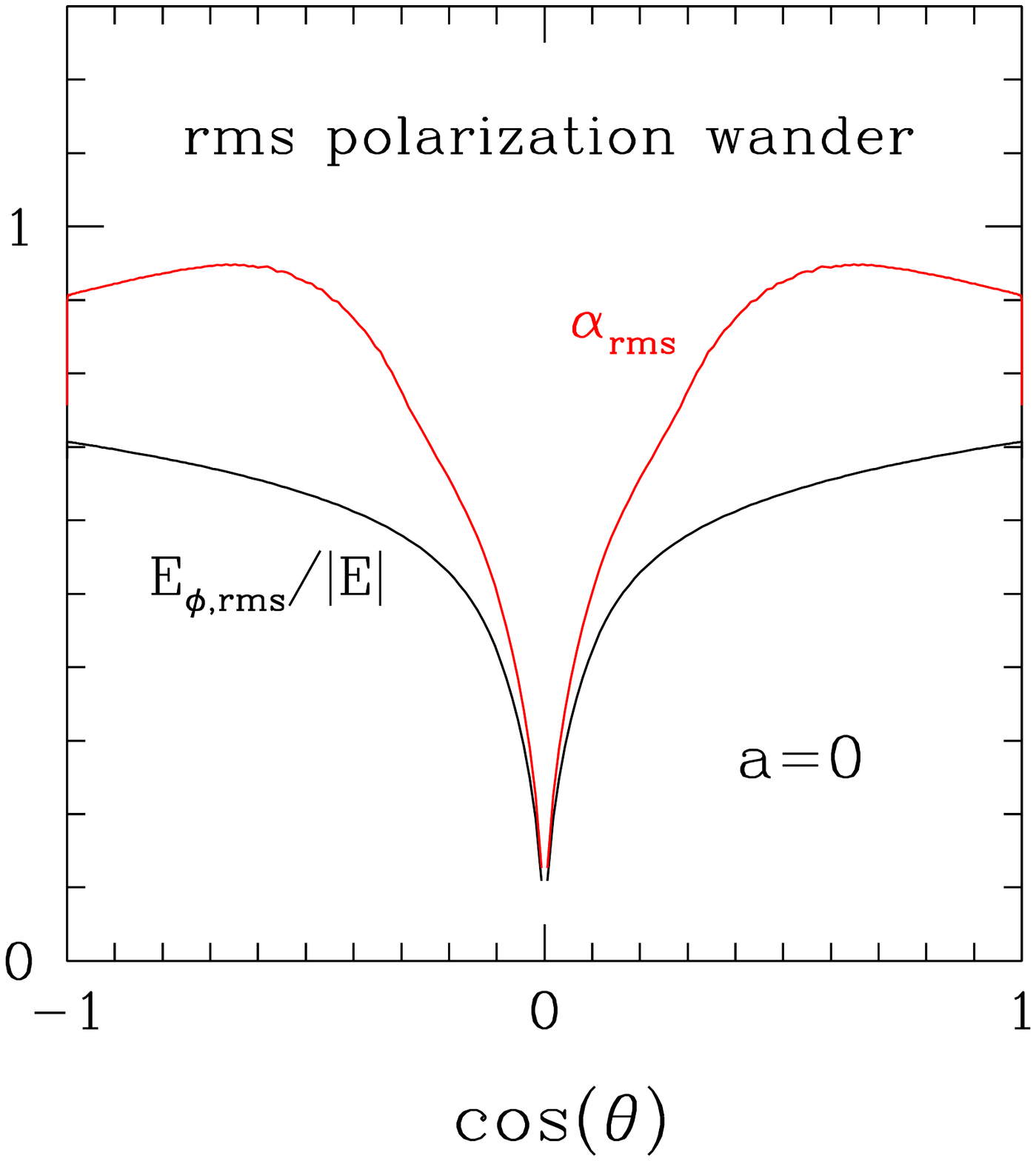}
\vskip -1in
\caption{Red curve: rms variation in the linear polarization angle of the prompt electromagnetic pulse, 
as seen by an observer positioned at a fixed latitude but random orbital phase relative
to the emission point.  Black curve:  rms $\phi$ component of the electric vector, also averaged over the
angle $\phi$ of emission.  Polarization variance is minimized where detection volume is maximized (Figure
\ref{fig:volume}).}
\vskip .2in
\label{fig:polarization_var}
\end{figure}

Observers with this favored orientation also see a more consistent polarization signature.  The emission process
we consider (upboosting of an ambient magnetic field by a tiny explosion) produces a high degree of linear
polarization when the damping length of the impulse is small compared with $r_{\rm isco}$.
Figure \ref{fig:polarization_var} shows the variance in the polarization angle, over successive releases
of energy at random orbital phases, as a function of the latitude of the observer.  This variance is minimized
just where the detection volume is maximized.  The implications of these results for the repeating burst emission of
FRB 121102 are discussed in Section \ref{s:frb}.

\subsection{Variation in the Explosion-BH Separation}

Although emission near the ISCO is a direct consequence of the model for FRBs that motivates our calculations,
it is worth briefly considering the possibility of larger separations between the emission radius $r_{\rm em}$ and the BH.  
For example, collisions between two compact stars are most likely to occur well outside the ISCO, and so a distant 
observer will have a smaller probability of seeing a strong gravitational lensing signature imposed on the gravity waveform.

The proportion of rays captured into the light ring, and subsequently released in caustics of secondary or higher degree,
would decrease as $\sim 1/r_{\rm em}^2$.   The brightness of the primary caustic, as measured at a fixed angular 
displacement $\alpha_{\rm obs} = |\theta_{\rm obs} - \pi/2|$ of the observer from the BH equator, is easily worked out
in the weak-gravity regime.  The fluence measured by an observer positioned at a distance $r$
is proportional to the magnification.  This is $\theta_E/2\theta_{\rm em-BH}$ (e.g. \citealt{weinberg2008}), where 
$\theta_E \simeq (4M r_{\rm em})^{1/2} / r$ is the angular radius of the Einstein ring that would be observed
at $\alpha_{\rm obs} = 0$, and $\theta_{\rm em-BH} \ll \theta_E$ is the angle between the direct lines from the emission
point and from the BH to the observer.   The latitude $\alpha_{\rm obs}$ of the observer is related to $\theta_{\rm em-BH}$ by $\alpha_{\rm obs} = (r/r_{\rm em})\theta_{\rm em-BH}$.  Hence the measured fluence is enhanced by a factor
$\alpha_{\rm obs}^{-1}(M / r_{\rm em})^{1/2}$, precisely the angular scaling that is measured in Figure \ref{fig:flux_lat}.
A randomly positioned observer has a probability $\sim M/4r_{\rm em}$ of seeing a strong lensing signature when
$r_{\rm em} \gg M$.

\section{Reflection off Quiescent, Cool Gas}\label{s:reflect}

Although radio telescopes are relatively sensitive to transient phenomena, as measured by the
electromagnetic energy received, the detection of low-frequency emission may be severely limited 
by absorption near the source \citep{pacholczyk}.  A SMBH must sustain a very low accretion rate for gigahertz waves to escape
from close to its horizon, even after allowing for the feedback of a strong electromagnetic
wave on ambient electrons (\citealt{thompson17b}).  If a thin disk component of the accretion flow is present,
the mass flow through it must also be substantially diminished.  However, in contrast with a hot and dilute 
accretion component, a cool disk would still present a barrier to the propagation of low-frequency waves
and an additional source of repetitions through reflection.  

The behavior of a quiescent disk is more complicated than the lensing effects we have investigated
so far, but some progress is possible in constraining the disk profile following 
an ionization transition.
If the mass transfer rate $\dot M_d$ through the disk drops far enough, the temperature at the disk midplane 
will reach $\sim 10^4$ K, below which the ionization level collapses.  Then the magnetic field decouples from
the disk material, mass transfer driven by the magnetorotational instability essentially halts, except
possibly in a thin surface layer \citep{gammie96}, and the disk mass profile freezes.  Such a cool, remnant
disk has been hypothesized to orbit the SMBH near the Galactic center \citep{ns03}.  In the extreme environment
of a SMBH, a surface layer should remain ionized by UV radiation and energetic particles, providing a
mirror for a strong electromagnetic wave.

This disk is light enough to remain gravitationally stable,  but outside a certain distance
from the SMBH, its self-gravity is strong enough for the disk to withstand warping by the 
differential rotation of the Kerr spacetime.  Close to the hole, the Lense-Thirring torque overwhelms
internal disk stresses, and the disk normal aligns with the BH spin \citep{bp75}.
Near the transition between these two regimes, which we now consider, the disk may break \citep{nealon15}.

The disk surface mass density $\Sigma$ is well constrained following the ionization transition.
The mass transfer rate toward a SMBH of mass $M_\bullet$ is expressed dimensionlessly as a fraction
$\dot m$ of the value that will source an Eddington-level dissipative luminosity in the inner disk, 
$\varepsilon_{\rm rad} \dot M_{\rm edd} c^2 = 4\pi GM_\bullet m_p c/\sigma_T$. (Here $\varepsilon_{\rm rad}$
is the radiative efficiency in the inner disk, and the other constants have their usual meanings.)  
The $r-\phi$ component of the internal stress (considered here at the onset of the ionization transition) is 
expressed in terms of a viscous stress feeding off the radial differential rotation and is normalized to a fraction
$\alpha$ of the gas pressure, $-\nu \rho R d\Omega/dR = \alpha P$.  Given that the mass transfer rate drops
gradually, it can be approximated by the steady-state formula
\citep{pringle81}
\be\label{eq:acc1}
\dot M_d = 3\pi \nu\Sigma = 2\pi \alpha \Omega h^2 \Sigma = {\dot m\over\varepsilon_{\rm rad}}
{4\pi GM_\bullet m_p\over \sigma_T c}
\ee
at a cylindrical radius $R \gg R_g \equiv GM_\bullet/c^2$.  The vertical scale height is expressed in terms of 
the orbital angular velocity, midplane temperature, and mean molecular weight as $h = \Omega^{-1}(\kB T_c/\mu)^{1/2}$.  

Vertical radiative energy transfer is mediated by absorption, with opacity near the hydrogen ionization threshold
\be\label{eq:kapp}
\kappa(P,T) = \kappa_0 T^{-\alpha} P^\beta.
\ee
Here, $\kappa$, temperature $T$ and $P$ are expressed in cgs units, and $\kappa_0 = 10^{12.026}$, $\alpha = 3.36$,
and $\beta = 0.928$ \citep{zhu09}.  The energy flux through each face of the disk (at a height $z \sim h$ above the
midplane) is
\be\label{eq:acc2}
{3\Omega^2\over 8\pi}\dot M = -{1\over\kappa\rho}{d\over dz}\left({4\over 3}\sigma_{\rm SB} T^4\right)
\sim {8\sigma_{\rm SB}T_c^4\over 3\kappa(T_c,P_c) \Sigma}.
\ee
The midplane gas pressure and disk surface density are related to the midplane density $\rho_c$ and scale height 
by $P_c = \rho_c (\Omega h)^2$ and $\Sigma \sim 2h\rho_c$.  Then Equations
(\ref{eq:acc1})-(\ref{eq:acc2}) yield two expressions for $\rho_c$ and $T_c$, which combine to give
\be
T_c(R) = 3\times 10^7\,(\alpha M_{\bullet, 6})^{-0.22}
\left({\dot m\over \varepsilon_{\rm rad}}\right)^{0.33} \left({R\over R_g}\right)^{-0.83}\quad{\rm K},
\ee
and 
\be\label{eq:sig}
\Sigma(R) = 2\times 10^6\,\alpha^{-0.78} M_{\bullet, 6}^{0.22}
\left({\dot m\over \varepsilon_{\rm rad}}\right)^{0.67} \left({R\over R_g}\right)^{-0.67}\quad{\rm g~cm^{-2}}.
\ee
Here the SMBH mass is normalized as $M_\bullet = M_{\bullet,6}\times 10^6\,M_\odot$.

The ionization transition ($T_c \leq 1\times 10^4$ K) occurs at an accretion rate
\be
{\dot m\over \varepsilon_{\rm rad}} = 1\times 10^{-3}\,(\alpha M_{\bullet, 6})^{0.67}\left({R\over 10^3\,R_g}\right)^{2.5},
\ee
when the surface density has dropped to 
\be\label{eq:sigR}
\Sigma(R) = 3\times 10^2\,\alpha^{-0.33} M_{\bullet, 6}^{0.67} \left({R\over 10^3\,R_g}\right)\quad{\rm g~cm^{-2}}.
\ee
Although such a cool disk is geometrically thin close to a SMBH, it is still gravitationally stable:  the Toomre parameter
\be
Q(R) \equiv {\Omega^2 h\over \pi G\Sigma} = 9\times 10^4 \alpha^{0.33}M_{\bullet,6}^{-1.67}\left({R\over 10^3\,R_g}\right)^{-5/2}
\ee
is well above unity inside $10^{4-5}$ gravitational radii, as long as the internal disk temperature remains
buffered at $10^4$ K.

The development of a warp in an orbiting thin disk, including the effects of self-gravity, viscous diffusion, and the Lense-Thirring 
torque, has been studied by \cite{td14}.  The self-gravitational torque by itself overcomes the Lense-Thirring torque outside
a radius
\be
R_w = \left[{\widetilde a_{\bullet} c^2 R_g^{5/2}\over \pi G \Sigma(R_w)}\right]^{2/7}.
\ee
Here $\widetilde a_\bullet = a_\bullet/R_g$ is angular momentum of the BH in units of $GM_\bullet^2/c$.
Because $\Sigma(R)/R$ is constant in the range of radius where Equation (\ref{eq:sig}) is valid, we have
\ba\label{eq:rw}
{R_w\over R_g} &\;=\;& \left[ {\widetilde a_\bullet c^2\over \pi G R_g^2 (\Sigma/R)}\right]^{2/9} \nn
               &\;=\;& 6\times 10^3\, \alpha^{2/27}\, \widetilde a_\bullet^{2/9} M_{\bullet,6}^{-10/27}.
\ea
We posit that the disk outside this radius is misaligned with the spin of the SMBH.  As the viscosity parameter $\alpha$ drops below 
$\sim 0.1 Q(R_w)h(R_w)/R_w$ (but is still large enough to suppress the propagation of bending waves, $\alpha \gg h(R_w)/R_w$), 
the equilibrium warp profile develops a strongly oscillatory component that migrates inward from $R \sim R_w$ (see Figure 8 of 
\citealt{td14}).  For a disk at the threshold of an ionization transition, this corresponds to 
$\alpha \lesssim 0.2 \widetilde a_\bullet^{-6/11}M_{\bullet,6}^{-25/22}$.  This oscillatory warp rises to a large amplitude a modest 
distance inside radius $R_w$.   The following discussion is based on the hypothesis that, at this stage, the 
disk fragments into thin annuli, an effect that is seen in numerical simulations of disks without self-gravity \citep{nealon15}.

\begin{figure}
\epsscale{0.9}
\plotone{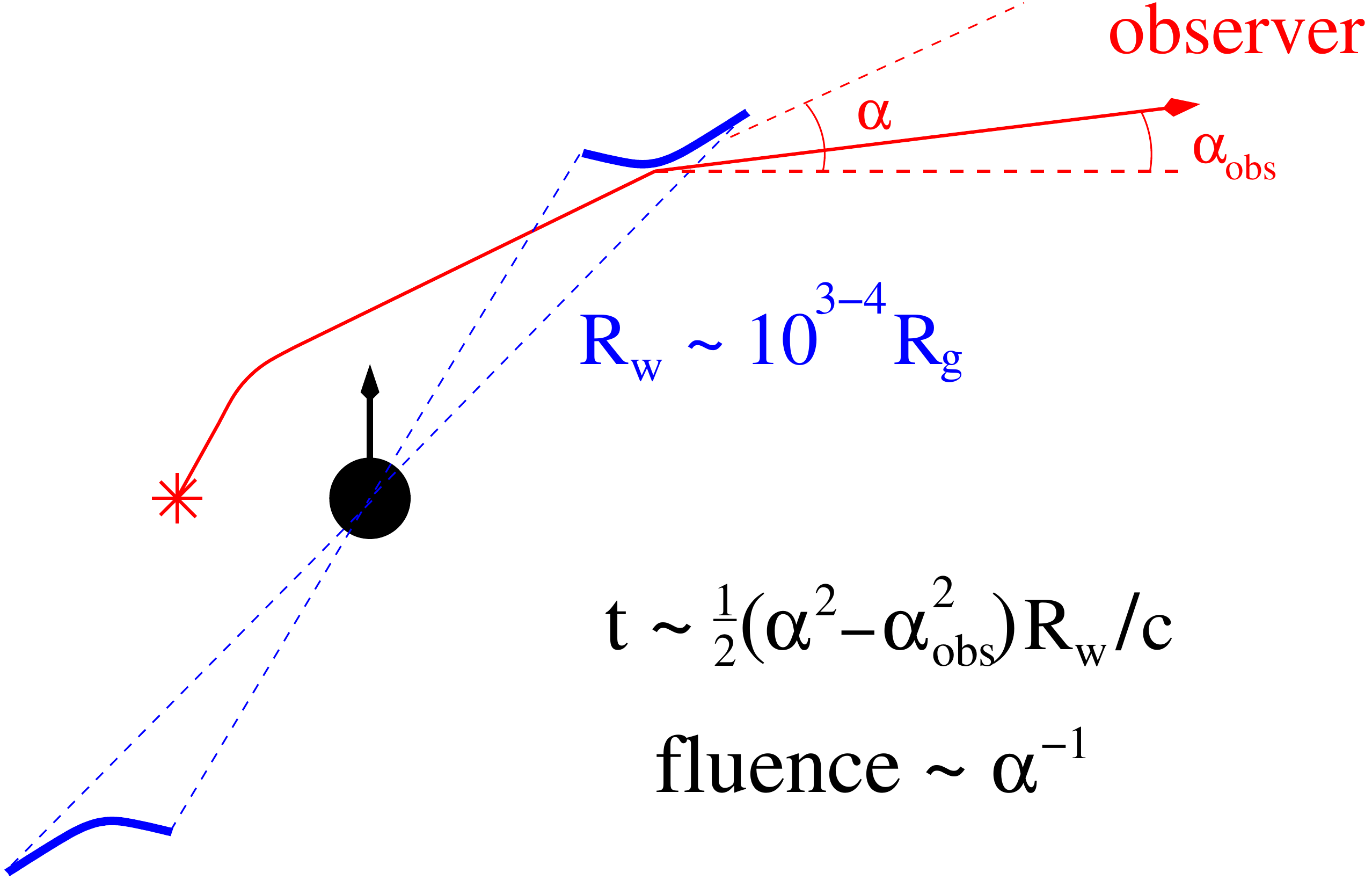}
\caption{Rays emitted by a tiny explosion near the ISCO of a SMBH converge into a gravitational caustic in a
direction nearly antipodal to the explosion site.  Rays propagating near the BH equator are also reflected
by annular fragments of a warped H/He disk concentrated around radius $R_w$ given by Equation (\ref{eq:rw}).
Reflection produces an additional geometrical delay depending on the angular offsets $\alpha_{\rm obs}$ and $\alpha$
from the BH equator of the observer and of the outgoing ray before reflection.  Fluence transported by the ray 
diverges as $\sim \alpha^{-1}$, and therefore decreases with geometrical delay (relative to unreflected rays) as 
$\sim t^{-1/2}$ for $t > \alpha_{\rm obs}^2R_m/2c$.}
\vskip .2in
\label{fig:reflect}
\end{figure}

\subsection{Reflection by Annular Fragments of a Warped Disk}\label{s:annulus}

Here we consider the reflection of a light pulse from a central source by fragments of a warped disk, which are 
represented by axisymmetric, tilted annuli. The emitted rays have already been gravitationally lensed\footnote{We
assume that the frozen disk, which would have a very low surface density near the ISCO (Equation (\ref{eq:sigR})), 
is present only at $R \gg R_g$.} by the SMBH;
we are considering here their subsequent interaction with cool H/He orbiting at a fairly large distance 
($R_w \sim 10^{3-4}R_g$) from the BH (Figure \ref{fig:reflect}).  The most strongly lensed rays travel closest to the
BH equator and also are most likely to interact with a warped disk that has aligned with the BH spin well inside
radius $R_w$.

The surface of each annulus is written in cylindrical coordinates $(R,\phi,z)$ aligned with the spin of the BH,
\be
z_d(R,\phi) = h(R)\sin\bigr[\phi-\phi_0(R)\bigl].
\ee
Here $\phi_0$ marks the nodal line.  The observer sits well beyond the reflection point 
in the cartesian direction $\hat x$.  The outgoing ray is nearly radial just before reflection, 
$\hat k = \hat r$, and after reflection has unit vector
\be
\hat k' = \hat k - 2(\hat k\cdot\hat n_d)\hat n_d,
\ee
where
\ba
\hat n_d &=& n_{d,0}\Bigl[\hat z - (\partial_R z_d)\hat R - R^{-1}(\partial_\phi z_d)\hat\phi \Bigr];\nn
   n_{d,0} &\equiv& [1 + (\partial_R z_d)^2 + R^{-2}(\partial_\phi z_d)^2]^{-1/2}
\ea
is the unit normal vector of the annulus at the reflection point 
$[R_{\rm ref}, \phi_{\rm ref}, z_d(R_{\rm ref},\phi_{\rm ref})]$.  Requiring $\hat k'$ to lie in the direction of
the observer leads to two constraints, the first on the magnitude of the radial disk warp at the reflection point
\be
   \left({R\over z_d}\partial_R z_d\right)_{R_{\rm ref},\phi_{\rm ref}} 
    = {1\over 2}\left[1- (\partial_Rz_d)^2 - R^{-2}(\partial_\phi z_d)^2\right]_{R_{\rm ref},\phi_{\rm ref}}
\ee
(from $\hat k'\cdot \hat z = 0$), and the second on the azimuthal angle of the reflection point
\be
\tan\phi_{\rm ref} = \left(-{z_d\partial_\phi z_d\over R(R + z_d\partial_Rz_d)}\right)_{R_{\rm ref},\phi_{\rm ref}}
\ee
(from $\hat k'\cdot\hat y = 0$).  The first expression states that the
radial warp  must be mild.  The second expression strongly limits the azimuthal offset of the reflection point from the
direct line of sight between the observer and the BH:  given that the nodal lines of the rings are randomly distributed in azimuth, one typically has $|\partial_\phi z_d| \sim |z_d|$.  (We focus here on the case where the outer disk is moderately
tilted from the equatorial plane of the BH, and not nearly perpendicular.)  Except for the strongest warps and longest
geometrical delays, the ray offset is small, $\phi_{\rm ref} \sim (z_d/R)_{\rm ref}^2$.  

There is now an interesting implication for the polarization of the brightest and most strongly lensed rays.
These rays reflect off disk annuli of height $z_d \ll R_{\rm ref}$, and the reflection point, as seen by a distant
observer positioned near the equator, lies nearly directly above (or below) the gravitational caustic:  
$\delta\phi/\delta\theta = \phi_{\rm ref} R_{\rm ref}/z_d \sim z_d/R_{\rm ref} \ll 1$.  In other words, the plane of 
reflection runs nearly in a longitudinal direction:  $\hat\theta\cdot (\hat k\times\hat k') \simeq 0$.  

In this situation, there is only a weak rotation of the polarization vector following reflection.  Treating the disk
surface as a conducting sheet, the unit electric vector after reflection is
\be
\hat E' = 2(\hat E\cdot\hat n_d)\hat n_d - \hat E,
\ee
the (irrelevant) minus sign deriving from the surface boundary condition ${\bm E}\times \hat n_d \rightarrow 0$.
Supposing the ray to be polarized $\hat\phi\cdot \hat E = 0$ before reflection (Figures \ref{fig:polarization_map},
\ref{fig:polarization_map2}), the electric vector after reflection has a small azimuthal component
\be\label{eq:pol2}
\hat E' \cdot \hat y \simeq -2\left({\partial_\phi z_d\over R}\right)_{R_{\rm ref},\phi_{\rm ref}}.
\ee
Gravitational lensing enhances the ray fluence in proportion to $\alpha^{-1} \sim R_{\rm ref}/z_d$;  hence, one
finds a small rotation of the polarization vector scaling inversely with the fluence of the reflected pulse.

\section{Application to Repeating FRBs}\label{s:frb}

\begin{figure}
\epsscale{0.8}
\vskip -0.3in
\plotone{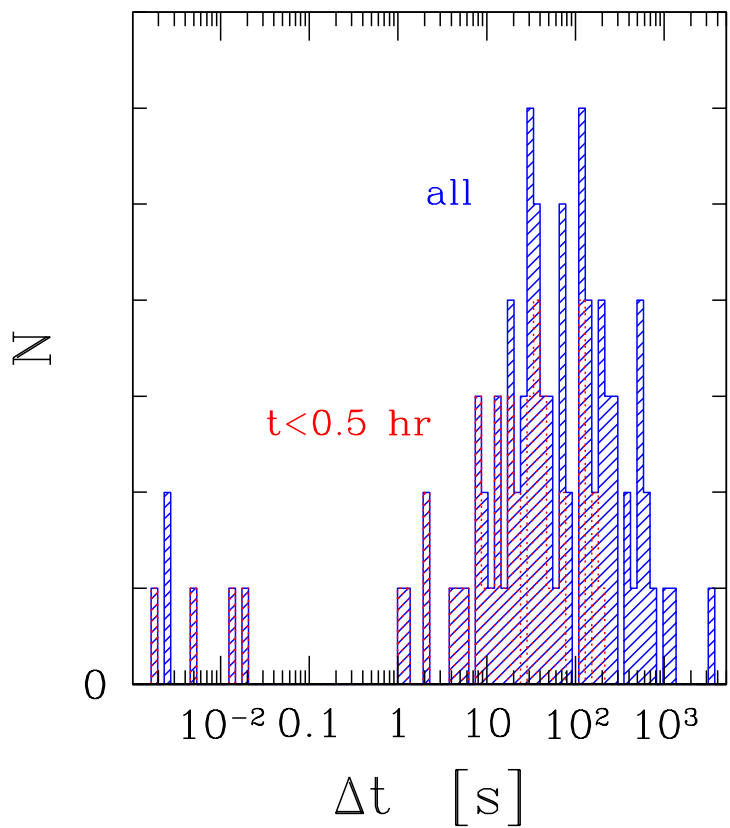}
\vskip -1in
\caption{Histogram of intervals between successive bursts, as recorded from FRB 121102 in a single $\sim 4.5$ hr cluster
by \cite{zhang18}.  Red histogram is the subsample of bursts arriving earlier than 0.5 hr from the start of the observation.}
\vskip .2in
\label{fig:interval}
\end{figure}

\begin{figure}
\epsscale{0.8}
\vskip -0.3in
\plotone{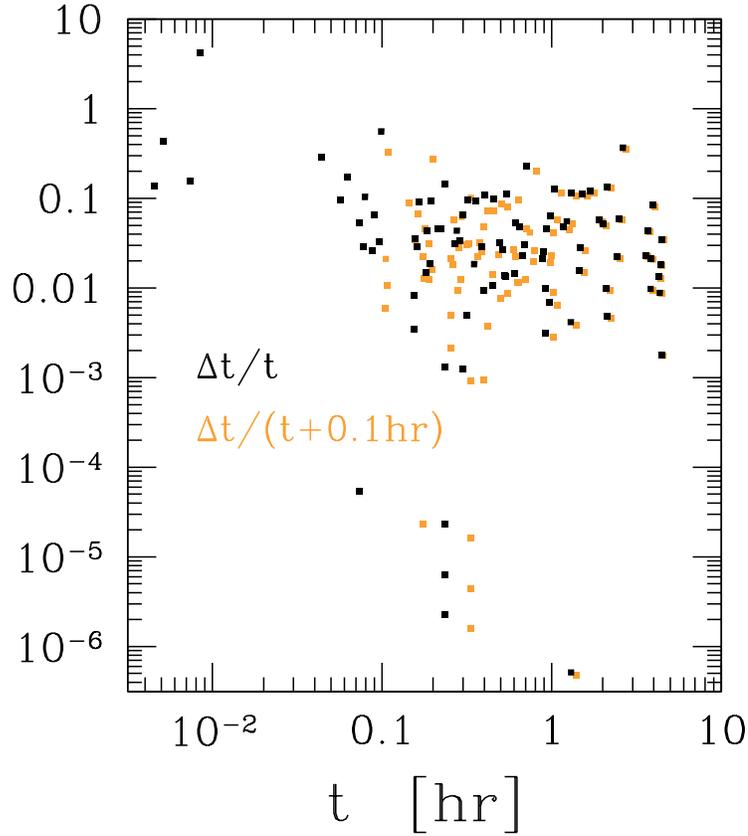}
\vskip -1in
\caption{Black points: intervals $\Delta t$ between successive bursts from FRB 121102, relative to the time $t$ 
since the start of observations as recorded by \cite{zhang18}.
The source was active at the start; hence, $\Delta t/t$ is artificially enhanced at small $t$.  Because the distribution
of $\Delta t/t$ flattens at large $t$, we may estimate the time that the source had been active before the start of
observations by adding an offset time to $t$.  The gold points show the result for an offset of $0.1$ hr.}
\vskip .2in
\label{fig:dt_TOA}
\end{figure}

\begin{figure}
\epsscale{0.8}
\vskip -0.3in
\plotone{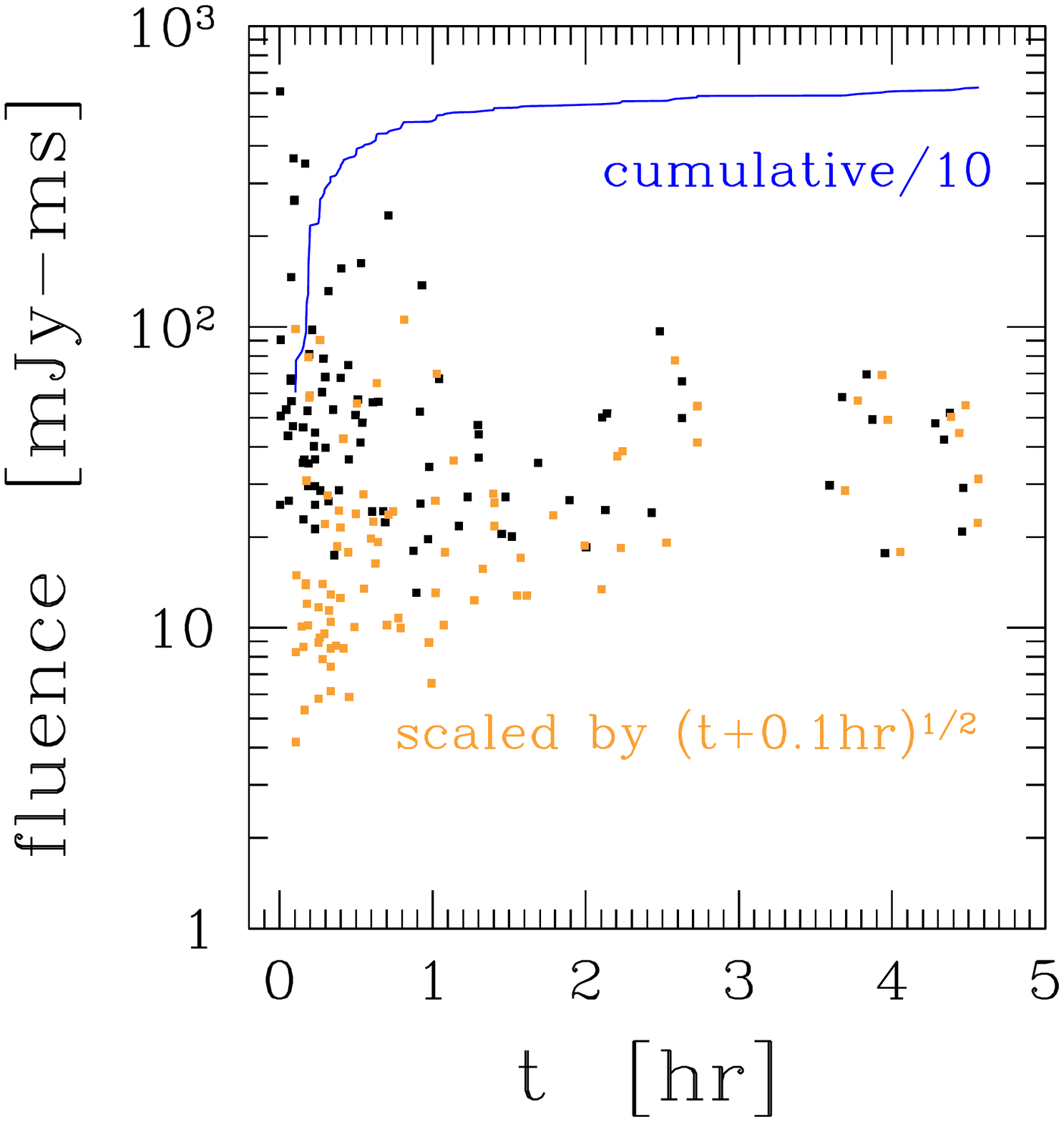}
\vskip -1in
\caption{Dependence of burst fluence on time since the onset of observations of FRB 121102 reported by \cite{zhang18}.
Gold points compensate for the scaling $F \propto t^{-1/2}$ expected from multiple reflections of a single 
gravitationally lensed impulse.  Gravitationally lensed burst fluence varies as $\sim \alpha^{-1}$ with angular offset 
$\alpha$ from the BH equator (Figure \ref{fig:flux_lat}), and the delay time as $\sim {1\over 2}\alpha^2 R_w/c$, where
$R_w \sim 10^4\,GM_\bullet/c^2$ is the distance of reflecting disk fragments from the BH (Equation (\ref{eq:rw})).}
\vskip .2in
\label{fig:fluence_TOA}
\end{figure}

A single FRB source, FRB 121102 \citep{spitler16}, is presently known to repeat.  Bursts from this source
sometimes arrive in clusters lasting a few hours, the best sampled of which was reported by \cite{gajjar18} and
\cite{zhang18} in a 5-8 GHz band.  Here we consider how a combination of gravitational lensing and reflection off annular
fragments of a quiescent disk may transform a single point emission near the ISCO of a SMBH into a burst cluster of a similar
duration \citep{thompson17a}.

\subsection{Effect of Gravitational Lensing}

The pattern of repetitions produced by gravitational lensing was worked out in
Sections \ref{s:bright}-\ref{s:time}, \ref{s:raych} and \ref{s:bias}.  It is possible, in principle, to measure
both the mass and spin of the BH by detecting both prograde and retrograde periodicities in the delayed pulses.  
Nonetheless, periodicity at fixed observer azimuth $\phi$ emerges only with a delay of $\sim (40-60)\,GM_\bullet/c^3$,
depending on the spin of the BH (Figure \ref{fig:tvsphi}), and is accompanied by exponential dimming (Figure
\ref{fig:fluxvsphi}).  In the case of a rapidly spinning BH, around which the prograde light rays dim relatively slowly,
the retrograde rays are greatly reduced in brightness compared with the prograde rays.  

The first repeat pulse arrives with a delay that varies smoothly with the azimuth of the observer relative to the emission
point.  In the case where observer and emission are nearly antipodal, a ray caustic is observed with a characteristic
delay $(5-10)GM_\bullet/c^3$ that depends on both $M_\bullet$ and $\widetilde a_\bullet = a_\bullet/R_g$ (Figure 
\ref{fig:fluxvst}).  The azimuth of peak brightening shifts significantly from the antipodal direction as 
$\widetilde a_\bullet \rightarrow 1$.  Given a favorable orientation relative to the BH equator, this delayed brightening
might be identified after monitoring multiple burst clusters, each sourced by an explosion at a random orbital phase.  

Figure \ref{fig:interval} shows the distribution of intervals between successive bursts in the FRB 121102 cluster reported by
\cite{zhang18}.  A few very rapid repetitions are observed, but then there is a large gap in repetition
time up to $\sim 10$ s, followed by a broad distribution of intervals extending up to $\sim 10^3$ s.  Identifying
the lower range of this broad distribution with a lensing delay of $(5-10)\,GM_\bullet/c^3$ implies a SMBH mass 
$M_\bullet = (2-4)\,\times 10^5\,M_\odot$.  Fast (approximately millisecond) repetitions could be produced self-consistently by the process
of specular reflection, during which the pulse intensity is high enough to induce trans-relativistic motion in the plasma
mirror \citep{thompson17a}.  The incidence of fast repetitions detected in FRBs also limits the abundance of an intervening
cosmological population of 30-100$\,M_\odot$ BHs through the effect of gravitational lensing \citep{munoz16}.

\subsection{Reflection of Lensed Rays off Fragments of a Quiescent Disk}

Although gravitational lensing may easily produce one or two bright repetitions, it cannot supply the large
number detected in the burst clusters of FRB 121102, or the longer burst intervals -- even if spin of the SBMH
is nearly extremal -- at least in the standard framework of Einsteinian relativity adopted here.
Most of the intervals the the range $10 < \Delta t < 30$ s are concentrated during the first 0.5 hr of the same observation,
whereas most of the longer intervals exceeding $\sim 300$ s are delayed to the later part (Figure \ref{fig:interval}).
The ratio $\Delta t/t$ between the interval to the next burst and the time of a given burst (recorded since the beginning
of the observation) is seen to have a relatively flat distribution over most of the observation (Figure \ref{fig:dt_TOA}). 
A rise in $\Delta t/t$ as $t \rightarrow 0$ is probably an artifact of the activation of the source at some time $t_0$
before the start of the observing window;  this feature can be compensated by setting $t \rightarrow t + t_0$, with
$t_0 \sim 0.1-0.3$ hr.

The flat distribution of $\Delta t/t$ is consistent with most of the repetitions being due to reflection off a large
number of cool H/He clouds, here modeled as annular fragments of a thin disk, which are uniformly distributed
in angle as seen from the BH.  Gigahertz waves can escape from the near-horizon region only if the electron density there is low,
and so any disk component of the accretion flow naturally has frozen out.   As discussed in Section \ref{s:reflect}, such
a quiescent disk may fragment at a distance $\sim R_w$ from the BH given by Equation (\ref{eq:rw}), as a result of a
competition between its self-gravity and the Lense-Thirring torque exerted by the BH.  

A ray directed at angle $\alpha = \theta-\pi/2$ from the BH equator, and then scattered into a direction 
$\alpha_{\rm obs} = \theta_{\rm obs}-\pi/2$ aligned with the observer, experiences a delay 
$t \simeq {1\over2}(\alpha^2-\alpha_{\rm obs}^2)R_w/c$ (Figure \ref{fig:reflect}).  The maximum delay $\sim R_w/c$ matches
that observed in the burst cluster of Figure \ref{fig:interval} ($\sim 4$ hr) for $M_\bullet \sim 4\times 10^5\,M_\odot$.  

\subsection{Lensing Bias and Dimming of Reflected Pulses}

FRBs emitted near the ISCO of a SMBH are predicted to be substantially brighter when the emission point is nearly
antipodal to the observer (Figures \ref{fig:flux_map} and \ref{fig:fluxvsphi}).  The detection volume of a repeating
FRB source is therefore increased if the the spin of the SMBH lies in the plane of the sky (Figure \ref{fig:volume}).  This
opens up an interesting mechanism for producing intermittent burst activity:  the source is harder to detect when the random
emission phase of the seed explosion is offset from the direction antipodal to the observer.   When the alignment is
favorable and the burst is strongly gravitationally lensed, one also expects a negative correlation between fluence and the additional delay time(s) produced by reflection, because the fluence scales with angular offset $\alpha$ of the outgoing ray
(before reflection) as $F \propto \alpha^{-1}$ (Figure \ref{fig:flux_lat}).  Hence
\be\label{eq:fluxt}
F(t) \propto \left({2ct\over R_m} + \alpha_{\rm obs}^2\right)^{-1/2},
\ee
where $\alpha_{\rm obs}$ is the offset of the observer from the BH equator.  Applying the inverse of this scaling to the burst
cluster of FRB 121102 significantly flattens the distribution of burst fluence with delay time (Figure \ref{fig:fluence_TOA}).

One also sees from Equation (\ref{eq:fluxt}) that lensing bias survives the addition of plasma reflection to the effects 
of gravitational lensing.  In fact, given the tendency of the inner disk to align with the spin of the BH, it is possible
that cool H/He blocks the direct line of sight between the Earth and the ISCO region of a SMBH associated with FRB 121102, so
that bursts are only detected in reflection.

\subsection{Linear Polarization}

Our final consideration is the polarization of the emitted bursts.  Strong linear polarization
is expected for tiny electromagnetic explosions emitted near the ISCO of a SMBH, because the damping length of
the injected subluminal pulse is much smaller than $r_{\rm isco}$ \citep{thompson17b}.  (By contrast, the damping
length is more than $\sim 10^{13}$ cm in interstellar conditions.)  This has been measured in three burst clusters of
FRB 121102, which also exhibit a uniform direction of source polarization \citep{michilli18}.

Rays propagating near the BH equator maintain a consistent polarization direction (Figure
\ref{fig:polarization_var}) as long as the magnetic field has a persistent (toroidal or poloidal) orientation near
the ISCO.  The same polarization direction is shared, for SMBHs of modest spin $\widetilde a_\bullet \lesssim 0.7$, 
by strongly lensed rays that propagate above or below the intensity cusp.  The reflection of these rays off annular fragments
of a quiescent disk largely preserves this polarization (Equation (\ref{eq:pol2})).  The alignment of the polarization is
predicted to weaken in pulses with increasing delay time (and decreasing pulse fluence).

A large Faraday rotation measure is a firm prediction of emission from close to a SMBH and has been recently observed
\citep{michilli18}.  The electron density near the SMBH that allows the escape of gigahertz waves is too small to sustain a
continuous jet that powers the persistent radio source reported by \cite{chatterjee17} and \cite{marcote17};  instead, 
the persistent emission must be powered by an outward-moving blast wave further from the hole.  Such a blast would 
suppress accretion onto the SMBH, as is needed for radio wave escape.
In this situation, the Faraday rotation measure could be dominated by material at the radius of the blast, which 
has been estimated to be $\sim 0.1-0.3$ pc in a fairly model-independent way \citep{beloborodov17,mm18}.  We note that 
this blast dimension is small enough to be contained inside the Bondi radius of a SMBH of mass $\sim 3\times 10^5\,M_\odot$,
meaning that the rotation measure will decrease with time as the material and
magnetic field swept up by the shock is spread over a larger volume; there is now tentative evidence for this
\citep{michilli18}.  For example, the blast radius increases with time as $R_s \sim t^{1/2}$ when the ambient
plasma density scales with distance $r$ from the SMBH as $n_e \propto r^{-1}$.  Taking the magnetic field to be the
corresponding equipartition value, $B(r) \propto r^{-1}$, the rotation measure scales as $RM(R_s) \propto R_s^{-1} 
\propto t^{-1/2}$.  

\subsection{Synopsis}

The same relative orientation of SMBH, emission point, and observer that results in (i) brightening of the emitted pulse
by strong gravitational lensing, (ii) enhanced detectability of the source, and (iii) a $\sim t^{-1/2}$ decay of the
fluence observed in independently reflected rays with different delay times, also produces (iv) a consistent direction
of linear polarization in the separate ray bundles.
We obtain two estimates of the mass of a SMBH host of FRB 121102; both point to $M_\bullet \sim 3\times 10^5\,M_\odot$. 
This is on the low end of the SMBH mass spectrum, as would be expected if the persistent radio counterpart of FRB 121102
\citep{chatterjee17,marcote17} is a BH:  this source resides in a dwarf galaxy with a low ($\lesssim 10^8\,M_\odot$) 
stellar mass \citep{tendulkar17}.

\section{Discussion}\label{s:summary}

We have calculated the time and polarization profiles of a pointlike electromagnetic explosion near a BH,
as seen by an observer at a large distance.
The underlying physical model involves a collision between two massive objects (magnetic dipoles in the
application to FRBs) orbiting near the ISCO.   Our main results are as follows.

1. The intensity and polarization profile on the sphere at infinity is calculated by ray tracing.  Gravitational
lensing produces a strong caustic peak in a direction nearly antipodal to the emission point (exactly for a
nonspinning BH), and is also accompanied by a broader Doppler peak representing the motion of the emission frame.
The fraction of the rays absorbed by the BH grows with increasing spin rate, as does the fraction of escaping rays that
arrive in the form of delayed pulses.

2. The time profile shows multiple impulses at each position $(\theta,\phi)$ on the sky, representing
rays that are emitted at different angles in the emission frame.  Measuring forward from the first detected
impulse, there is initially no periodicity.  Two distinct periods eventually emerge, representing the
orbital periods of the prograde and retrograde light rings.

3. The delayed pulses produced by prograde and retrograde rays connect with each other and with polar
rays at discrete azimuths.  Caustics coincide with the connection points between polar and equatorial rays.
The azimuth of a given caustic shifts relative to the preceding caustic of the same type (except in the case
of vanishing BH spin).  The decay of the fluence transmitted to a fixed observer by successive pulses 
agrees with the fundamental spin-1 quasi-normal mode in the Schwarzschild case, but such a comparison is
complicated by the aperiodic intervention of caustic features when the BH spins.

4. We examine the polarization profile produced by emission due to the upboosting of an ambient magnetic field
into a propagating superluminal transverse mode by a small explosion.  The electric 
vector measured by an equatorial observer at a large distance from the BH maintains a fairly uniform orientation 
over a wide range of azimuth relative to the emission point.

5. There is an observational bias in favor of observing {\it repeated} electromagnetic explosions from
a BH whose spin is oriented in the plane of the sky.

As for the application to FRB 121102 and other potential repeating FRB sources, we note the following.

6. Strong lensing of electromagnetic bursts emitted at random phases near the ISCO is a possible source of apparent
intermittency: only if the emission point is nearly antipodal with the direction of the observer is the lensing 
amplification strong.  Intermittent growth and decay of the accretion rate onto the hole, associated with rising and
falling synchrotron absorption, could also sporadically shield gigahertz waves from detection \citep{thompson17a}.
These two possibilities can be distinguished by coordinated measurements at high and low frequencies.

7. Strong linear polarization is predicted for repeating FRBs, because of the short damping length of the 
injected energy compared with the radius of the ISCO and, therefore, compared with any plausible coherence length 
of the ambient magnetic field.  This polarization will maintain a consistent direction if the observer
lies in the equatorial plane of the SMBH, as lensing bias would suggest he or she is most likely to.
It will also be maintained by plasma reflection when the burst is strongly gravitationally lensed -- that is,
if the emission point and the direction of the observer are nearly antipodal.  A test of our approach
is that any repeating FRB source must have a large Faraday rotation, a general property of the plasma around SMBHs;
and, conversely, FRBs with modest measured rotation measures must not repeat.

8. Repeat bursts produced by reflection off annular fragments of a quiescent disk orbiting the SMBH around the radius
given by Equation (\ref{eq:rw}) will have a peak flux decaying as $\sim t^{-1/2}$ with the time 
since the first detected burst, once again if the emission point and the observer are nearly antipodal.  A burst
cluster from FRB 121102 reported by \cite{gajjar18} and \cite{zhang18} is consistent with this, and also shows
a flat distribution of fractional time intervals $\Delta t/t$ between bursts.  

9. More generally, the detection of repeating FRB pulses emitted from near-horizon regions of SMBHs would offer valuable
probes of plasma dynamics in strong gravitational fields and would constrain departures from Einsteinian gravity
or the presence of bound states of relativistic fields, such as light axions.

\acknowledgements
The author would like to thank Aaron Zimmerman for conversations, and the NSERC of Canada for financial support.



\vfil\eject

\appendix

\section{Computational Procedures}

We describe some details of the computation and numerical tests.

\subsection{Tetrad for a Circular Orbit}

A photon is emitted with 4-momentum $k_{\rm em}^a$ as measured in the rest frame of a massive orbiting
particle, labeled `em', with angular velocity given by Equation (\ref{eq:omegac}).   
This frame is connected by a Lorentz boost with the ZAMO frame rotating with angular velocity (\ref{eq:omega}).
This boost mixes the (0) and (3) tetrad components of the momentum, while preserving the (1) and (2) components 
(those parallel to $\hat r$ and $\hat\theta$ in B-L coordinates).  Hence
\ba\label{eq:pz1}
k^{(0)}_{\rm zamo} &=& \gamma_{\rm em}\left(k^{(0)}_{\rm em} + \beta_{\rm em} k^{(3)}_{\rm em}\right);  
\quad\quad k^{(1,2)}_{\rm zamo} = k^{(1,2)}_{\rm em};  \nn
k^{(3)}_{\rm zamo} &=& \gamma_{\rm em}\left(k^{(3)}_{\rm em} + \beta_{\rm em} k^{(0)}_{\rm em}\right),
\ea
where $\beta_{\rm em}$ is given by Equation (\ref{eq:betadef}).  The 4-momentum in the ZAMO frame is constructed 
from the ray tangent vector measured in the B-L frame as
\ba\label{eq:pz2}
k^{(0)}_{\rm zamo} &=& (-\widetilde g_{tt})^{1/2}{dt\over d\lambda};\nn
k^{(3)}_{\rm zamo} &=& g_{\phi\phi}^{1/2} \left({d\phi\over d\lambda} - \omega{dt\over d\lambda}\right); \nn
k^{(1)}_{\rm zamo} &=& g_{rr}^{1/2} {dr\over d\lambda} = k^{(1)}_{\rm em}; \quad\quad k^{(2)}_{\rm zamo} = 
g_{\theta\theta}^{1/2}{d\theta\over d\lambda} = k^{(2)}_{\rm em}.
\ea
Equating expressions (\ref{eq:pz1}) and (\ref{eq:pz2}) gives the tetrad (\ref{eq:tet}).

\subsection{Christoffel Symbols and Riemann Tensor for Kerr Spacetime}

The Christoffel symbols corresponding to the metric (\ref{eq:kerr}) are tabulated in Equation (2.14) of \cite{mg09}.
The Riemann tensor is constructed using the \cite{carter73} tetrad, $R_{\mu\nu\rho\sigma} = 
e^a_\mu e^b_\nu e^c_\rho e^d_\sigma R_{abcd}$, where
\ba
e^{(0)}_\mu dx^\mu &=& \left({\Delta\over\Sigma}\right)^{1/2}(dt - a\sin^2\theta d\phi); \quad\quad 
e^{(1)}_\mu dx^\mu = \left({\Sigma\over\Delta}\right)^{1/2}dr; \nn
e^{(2)}_\mu dx^\mu &=& \Sigma^{1/2} d\theta; \quad\quad  
e^{(3)}_\mu dx^\mu = {\sin\theta\over \Sigma^{1/2}}\bigl[adt - (r^2+a^2)d\phi\bigr],
\ea
and the tetrad components of the Riemann tensor are
\ba
R_{0101} &=& -R_{2323} = -2R_1;\quad\quad R_{0202} = R_{0303} = -R_{1212} = -R_{1313} = R_1;\nn
R_{0123} &=& 2R_2; \quad\quad R_{0213} = -R_{0312} = R_2.
\ea
Here
\ba
R_1 &\equiv& {Mr\over\Sigma^3}(r^2 - 3a^2\cos^2\theta);\nn
R_2 &\equiv& - {a M\cos\theta\over\Sigma^3} (3r^2 - a^2\cos^2\theta).
\ea
This construction has been tested by computing $R_{\mu\nu\rho\sigma}$ directly from the Christoffel symbols.

\subsection{Testing Solutions to the Geodesic Equation}\label{s:test}

Two numerical tests of the correctness of the geodesic solution are easily available.  The tangent
vector must satisfy the null equation $g_{\mu\nu}(dx^\mu/d\lambda)(dx^\nu/d\lambda) = 0$; 
and a ray trajectory without radial or angular turning points is easily compared with the solution 
obtained by direct quadrature, using the integrals ${\cal E}$, ${\cal L}_z$ combined with the Carter constant
${\cal Q}$.   For a ray originating on the equatorial plane,
\be
{\cal Q} = g_{\theta\theta} \left({d\theta\over d\lambda}\right)^2.
\ee
The $t$ and $\phi$ components of the tangent vector are obtained from the integrals
(\ref{eq:integrals}), and the other components satisfy the equations \citep{fn98}
\ba
\Sigma {dr\over d\lambda} &\;=\;& \pm R(r)^{1/2}; \quad\quad R \equiv \bigl[{\cal E}(r^2+a^2)-{\cal L}_z a\bigr]^2 -
       \Delta\,\bigl[ {\cal Q} + ({\cal L}_z - a{\cal E})^2\bigr];\nn
\Sigma {d\theta\over d\lambda} &\;=\;& \pm \Theta(\theta)^{1/2}; \quad\quad \Theta 
\equiv {\cal Q} - \cos^2\theta \left({{\cal L}_z^2\over\sin^2\theta} - a^2{\cal E}^2\right).
\ea

\section{Comparing Monte Carlo Method with Solution to Focusing Equation}\label{s:raychcomp}

\begin{figure}
\epsscale{0.9}
\plotone{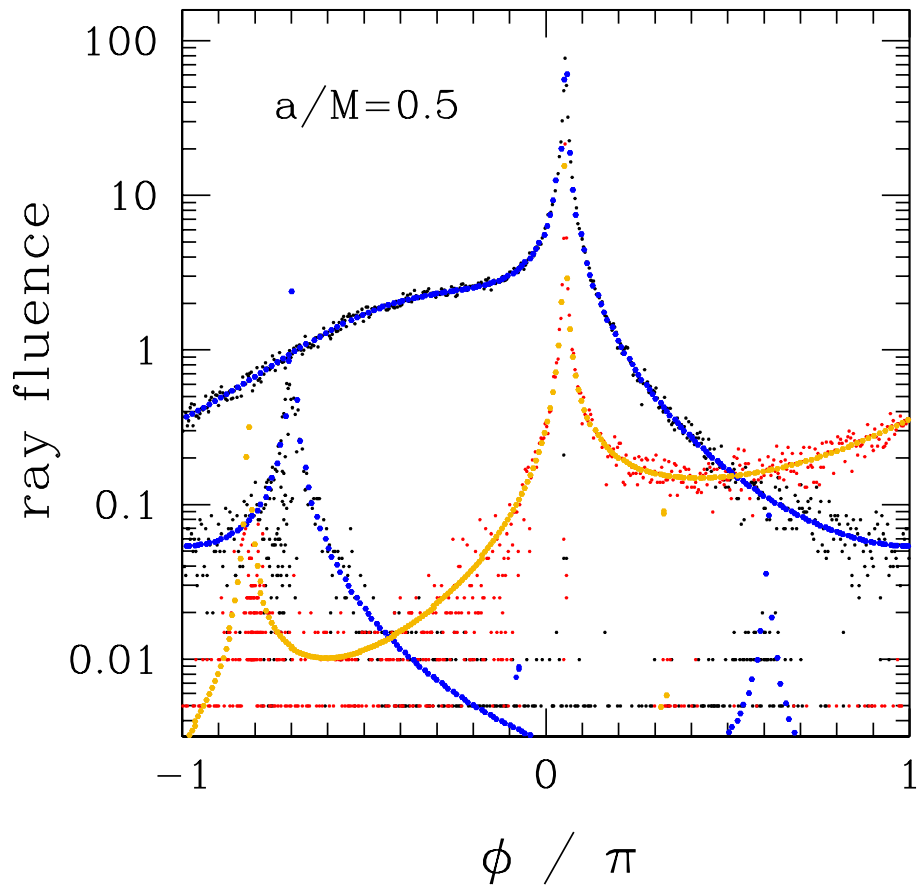}
\caption{Large points (blue and gold):  solution to the Raychaudhuri equations (\ref{eq:Raych2})
for a BH with spin $a/M = 0.5$, as a function of the azimuth of the ray at a large distance from the BH.  
Small scattered points (black and red):  ray fluence as calculated by a Monte Carlo technique with $2^{23}$ rays and
angular cells of size $\Delta\theta = 2^{-8}\pi$ and $\Delta\phi = 2^{-8}(\sin\theta)^{-1}\pi$.  Black and blue points:
prograde photon orbits; red and gold points: retrograde orbits.}

\vskip .2in
\label{fig:raychtest}
\end{figure}

Two methods can be compared for evaluating the ray fluence on the rotational equator of the BH, projected onto
the sphere at infinity:  first, a straightforward Monte Carlo technique and, second, an evaluation of the ray expansion
starting from a small sphere surrounding the emission point, using the Raychaudhuri equations (\ref{eq:Raych2}).
Figure \ref{fig:raychtest} shows excellent agreement, thereby supplying a nontrivial test of both methods. 
It should also be noted that, although the ray expansion is calculated for purely equatorial orbits, the 
caustics seen in Figure \ref{fig:raychtest} can be captured by the Monte Carlo technique only by including
nonequatorial orbits.  The method of Section \ref{s:raych}, which integrates through singularities of the 
focusing equation by combining the expansion with the one nonvanishing component of the ray shear tensor, 
is specialized to equatorial orbits.

\end{document}